%% file: arxiv_version.tex
\documentclass{aa}

\usepackage{graphicx}
\usepackage{txfonts}
\usepackage{float}
\usepackage{gensymb}
\usepackage{subfigure}
\usepackage{orcidlink}
\usepackage{comment}
\usepackage{multirow}
\usepackage{booktabs}
\usepackage{soul}
\usepackage{comment}
\usepackage{textcomp} 
\usepackage{pifont}   
\usepackage{booktabs}
\usepackage{amssymb}
\usepackage{amsmath}
\DeclareUnicodeCharacter{2212}{-}

\usepackage{hyperref}

\usepackage{amsfonts}

\usepackage[normalem]{ulem}
\usepackage[x11names]{xcolor}
\usepackage{xcolor}

\newcommand{\FIG}[1]{#1}

\begin{document}

   \title{Secondary small-scale dynamics of a Rayleigh-Taylor unstable solar prominence}

\author{M. Changmai \inst{\ref{KUL}}\orcidlink{0000-0002-3356-2398}
            \and 
            J. M. Jenkins\inst{\ref{ESAC},\ref{KUL}}\orcidlink{0000-0002-8975-812X}
            \and 
            R. Keppens\inst{\ref{KUL}}\orcidlink{0000-0003-3544-2733}
            }

\institute{Centre for mathematical Plasma Astrophysics, Department of Mathematics, Celestijnenlaan 200B, 3001 Leuven, KU Leuven, Belgium \label{KUL}\\
\and 
European Space Agency (ESA), European Space Astronomy Centre (ESAC), Camino Bajo del Castillo, s/n, Villanueva de la Cañada, Madrid, 28692, Spain \label{ESAC}
}

  \abstract
  % context 
  {}
  % Aims
{Quiescent solar prominences clearly show small-scale dynamics in observations. Their internal properties realize density contrasts with the Sun's atmosphere that must be liable to mainly Rayleigh-Taylor (RT) instabilities, which in turn lead to the formation of a vertically dominated prominence structure when viewed at the solar limb. As a result, prominences develop bubbles and plumes but also secondary instabilities in the form of Kelvin-Helmholtz (KH) roll-ups along the edges of the bubbles and plumes. Recent observations also indicate the existence of reconnection events within the Rayleigh-Taylor turbulent flows.}
% Methods
{We run 2.5D high resolution resistive magnetohydrodynamic simulations with the open-source {\tt MPI-AMRVAC} code. We achieve a spatial resolution of $\sim 11.7$~km in a 2D domain of size 30~Mm~$\times$~30~Mm and run the simulation for about 10 minutes of solar time. A dense, magnetic pressure supported prominence at coronal heights acts as our initial state which is then perturbed by the RT instability at the prominence corona interface. The localized current concentrations which are formed as a result of multiple, interacting instabilities are studied statistically.}
%Results
{The combination of primary (RT) and secondary (KH) instabilities leads to the formation of current sheets and causes an interplay between turbulent flow patterns and magnetic reconnection. The outflows from the reconnection regions lead to the formation of energetic jets from the current sheets facilitating energy exchange and dissipation throughout the prominence structure. We analyze our high-resolution prominence simulation using synthetic images of the broadband SDO/AIA 094, 171, and 193~\AA~and narrowband H$\alpha$ filters to compare the developing fine-scale structures with their observational counterparts. We find the majority of the secondary instabilities to form in the hotter regions surrounding the cooler prominence material.}
  % conclusions 
  {The secondary instabilities and current sheets in our simulation agree with observations (scale, speed, duration) yet the simulated activity localises predominantly in hot, surrounding coronal plasma rather then the cool prominence material. This inconsistency points to missing ingredients rather than a shortcoming of the KH interpretation, and so 3D follow-up studies should revisit these findings in more realistic magnetic topologies.}


\titlerunning{Small-scale prominence dynamics}

   \maketitle
%
%-------------------------------------------------------------------

\section{Introduction} \label{s:introduction}

Quiescent prominences are large, cool, and dense structures of plasma that exist stably in the hot solar corona. They have their origins in filament channels and manifest themselves above magnetic polarity inversion lines (PIL), which are the demarcation between mostly positive and negative magnetic fields at photospheric heights. These prominences are typically observed in quiet areas within the Sun's atmosphere, spanning various latitudes and occupying the most extensive scales which range from 60-600~Mm in length, 15-100~Mm in height, and 5-15~Mm in thickness \citep{1995ASSL..199.....T}. Early studies of their fundamental properties deduced characteristic temperatures and densities of $\sim 10^4$~K, densities $\sim 10^{−13}$~g~cm$^{-3}$, respectively, with initial magnetic field strength estimates around 3-10~G \citep{1995ASSL..199.....T}. Such ranges insist that prominences contain dynamics that are low plasma $\beta$ in nature, embedded within the equivalently low plasma beta solar corona. Indeed, \citet{1967ApJ...150..313R} calculated prominence $\beta$ values of order $\approx$ 0.02, as derived from average measurements of the magnetic field $\approx$ 10~G, and the pressure $\approx$ 0.1~dyne~cm$^{-2}$. In more recent observations, \citet{2003ApJ...598L..67C} have reported the existence of localised regions within quiescent prominences that exhibit significantly higher magnetic field strengths, ranging from 60-80~G.

Early studies of prominences focused on the velocity field, in particular noting the presence of upflows and downflows on the order of a few km~s~$^{-1}$ \citep[][]{engvold1981small,kubota1986vertical}. \citet{stellmacher1973} reported the development of a cavity within a quiescent prominence, as seen in H$\alpha$ slit-yaw-images, and linked it with a fast magnetohydrodynamic (MHD) wave signal. \citet{1981SoPh...70..315E} initially interpreted observed downflows as occurring along magnetic field lines, that are driven by the force of gravity. Motions observed in prominences include the formation of shear flows, strongly suggesting the presence of instabilities arising from this shear. Indeed, \citet{liggett1984rotation} observed the presence of rotating regions within non-eruptive prominences (5 out of 51 observed cases), with sizes between $3000\times 7000 \, \mathrm{km}^{2}$ and up to $30\times 104 \,\mathrm{Mm}^2$ and exhibiting rotation rates of around 30~km~s$^{-1}$. Shortly thereafter, however, early observations of magnetic fields in prominences revealed a predominantly horizontal magnetic field within the developing vertical plasma structure \citep[][]{1989ASSL..150...77L}. On first inspection, it is not at all clear how shear flows can develop perpendicular to the magnetic field under low-beta conditions - in 2D for a magnetic field oriented across the plane this is prohibited without significant fieldline deformation, thus in clear contradiction with the observations.

The high-resolution \textit{Hinode} \textit{Solar Optical Telescope} (Hinode; \citealp{Kosugi2007}, SOT; \citealp{Tsuneta2008}) has improved our understanding of the non-linear behavior of prominence internal dynamics \citep[][]{chae2010dynamics,hillier2012numerical}. \citet{chae2010dynamics} found from the observations that the vertical fine substructures, termed knots, were impulsively accelerated in the downward direction. The authors related this to magnetic reconnection and the interchange of the magnetic configuration. The SOT observations of \citet[][]{Berger_2008,Berger_2010,berger2011magneto} simultaneously linked these dynamics to the Rayleigh-Taylor (RT) instability \citep[][]{hillier2012numerical, hillier2012numerical2,keppens2015solar,xia2016formation}; in a 3D volume, or a carefully constructed 2D plane, such cross-field-like behaviour is permitted even in low-beta plasmas. In turn, the Kelvin-Helmholtz (KH) instability is a well-known shear flow instability that leads to the fragmentation of coherent vorticity structures into individual vortices during its non-linear development. In relation to solar prominences, {and in tandem with} the RT instability, the KH instability has subsequently been investigated in numerical works where good qualitative and quantitative agreement has been found with the observations \citep[][]{2017ApJ...850...60B, 2018ApJ...864L..10H}. \cite{Ryutova2010} combined SOT observations with theoretical criteria to categorize observed plasma instabilities in quiescent prominences, identifying RT and thermal instability (TI) processes, as well as how prominence cavities form.

Summarizing, quiescent prominences are seen to exhibit spatio-temporal evolution characterized by high variability. It is particularly interesting that the dynamic nature and the large prevailing Reynolds numbers then indicate the fluctuations present within prominences to be turbulent in nature. For example, \citet{2017ApJ...850...60B} used the SOT to study and analyse the characteristics of a quiescent prominence bubble and the associated instability dynamics. The characteristic speed of the largest rising bubble was found to be 1.3~km~s~$^{-1}$. The prominence downflows deposited plasma onto the bubble, forming an increasingly thick boundary layer with a speed of 20-35~km~s~$^{-1}$. This led to a strong shear flow of the order of 100~km~s~$^{-1}$ across the bubble boundary and in turn the coupled KH-RT instability. The nonlinear evolution of the KH instability is known to play a significant role in the generation of turbulent flows by means of reconnection and secondary instabilities \citep{2004GeoRL..31.2807M}.

Solar prominences are dictated by complex interactions between gravitational, thermal, and magnetic forces. In \citet{2023A&A...672A.152C} (hereafter MC23), we modelled this complex evolution as a Rayleigh-Taylor (RT) instability that later developed fully turbulent characteristics under the ideal-MHD approximation. We performed high-resolution simulations where a stratified atmosphere containing a prominence structure evolves due to Rayleigh-Taylor dynamics. This RT initiated at the density inversion at the bottom of the prominence, eventually deforming the prominence body into falling fingers and rising bubbles. Our study addressed the far-nonlinear turbulent state of the entire prominence body, as fingers got reflected upwards at the transition region and interacted with still-falling prominence matter \citep[cf.][]{Rees-Crockford:2024}. The truly mixed and turbulent state obtained was analyzed statistically, performing structure-function analysis to quantify the degree of intermittency. A fair agreement with a purely observational study on observed RT turbulence in prominences by \citet{leonardis2012turbulent} was obtained. Recently, \cite{2021A&A...651A..60H} used \textit{Interface Region Imaging Spectrograph} (IRIS; \citealp{DePontieu2014}) slit-jaw images at Si IV and Mg II lines, finding evidence for bi-directional jets that would signal internal magnetic reconnections within quiescent prominence bodies. They interpret these reconnections as resulting from shearing, buoyant flows. In addition to the general theory, we use this observation to motivate this follow-up numerical study, where we will investigate secondary instability processes in particular, within RT deforming prominences. We run highly resolved, resistive MHD models to focus on secondary shear flow effects, KH vortices, and induced reconnections.

This manuscript is organised as follows: we talk about the numerical setup in §~\ref{s:numerical_setup}, and in §~\ref{s:results} we analyze and characterize the energetic events due to the secondary instabilities. In §~\ref{s:discussion} and~\ref{s:conclusions}, we present our summary and conclusions, respectively.

\section{Numerical setup}\label{s:numerical_setup}

To investigate small-scale dynamics in a quiescent prominence due to initial RT instability, we performed 2.5D high-resolution resistive MHD simulations of a prominence inserted in the solar atmosphere using the parallelized, open-source Adaptive Mesh Refinement Versatile Advection Code \citep[{\tt MPI-AMRVAC};][]{keppens2012parallel,porth2014mpi,xia2018mpi,keppens2020mpi,keppens2023amrvac}. This work employs the same methodology as in MC23. Using Cartesian geometry, we establish a large simulation domain encompassing $30$Mm horizontally along the solar surface and $30$Mm vertically from the photosphere \,--\, low chromosphere into the solar corona. We use a maximum refinement level of seven and a base grid resolution of $40 \times 40$ to obtain a final grid resolution of $2560 \times 2560$. We achieve a spatial resolution of $\sim 11.7$~km and run the simulation for about 10 minutes of solar time. Compared to MC23, we added another AMR level to achieve this higher resolution (MC23 had cells of 23 km size), and we add explicitly resolved resistivity.

We solved the resistive MHD equations for the conservation of mass, momentum, total energy density, and induction. The equations are given by
\begin{eqnarray}
\label{eqn:mhd1}
        \partial_t \rho + \nabla \cdot (\textbf{v} \rho) = 0 \,,\\
\label{eqn:mhd2}   \partial_t(\rho \textbf{v}) + \nabla \cdot (\textbf{v}\rho \textbf{v} - \textbf{BB}) + \nabla p_{tot} = \rho \textbf{g} \,,\\
\label{eqn:mhd3}   \partial_t e + \nabla \cdot (\textbf{v}e - \textbf{BB}\cdot \textbf{v} + \textbf{v}p_{tot}) = \rho \textbf{g}\cdot\textbf{v} + \nabla \cdot (\textbf{B} \times \eta \textbf{J})\,,\\
\label{eqn:mhd4}   \partial_t \textbf{B} + \nabla \cdot (\textbf{vB} - \textbf{Bv}) = - \nabla \times (\eta \textbf{J})\,,
\end{eqnarray}
where $p = (\gamma - 1)(e - \rho \textbf{v}^2/2 - \textbf{B}^2/2)$, $p_{tot} = p + \textbf{B}^2/2$, and $\textbf{J} = \nabla \times \textbf{B}$. The quantities $\rho$, $\textbf{v}$, $p$, $p_{tot}$, $e$, and $\textbf{B}$ denote density, velocity vector, plasma pressure, total pressure, total energy density, and magnetic field vector, respectively, and $\gamma$ (adiabatic index) is the ratio of specific heats, taken as 5/3 under the assumption of a mono-atomic ideal gas.  A uniform resistivity,
$\eta = 10^{-4}$ (or $1.2 \times 10^{13}$~cm$^2$~s$^{−1}$ in physical units) is taken throughout the entire simulation domain. The magnetic field vector is measured in units for which magnetic permeability is equal to $\mu_0 = 1$. The plasma temperature, $T$, follows from the ideal gas law,
\begin{equation}
    p \mu = R \rho T \,,
\end{equation}
where $R$ and $\mu$ are the gas constant and mean molecular mass, respectively. The normalization of the quantities is done identically to MC23.

\begin{figure*}[]
\centering
        \resizebox{0.9\hsize}{!}{
        \includegraphics[clip=, trim= 0 0 0 0]{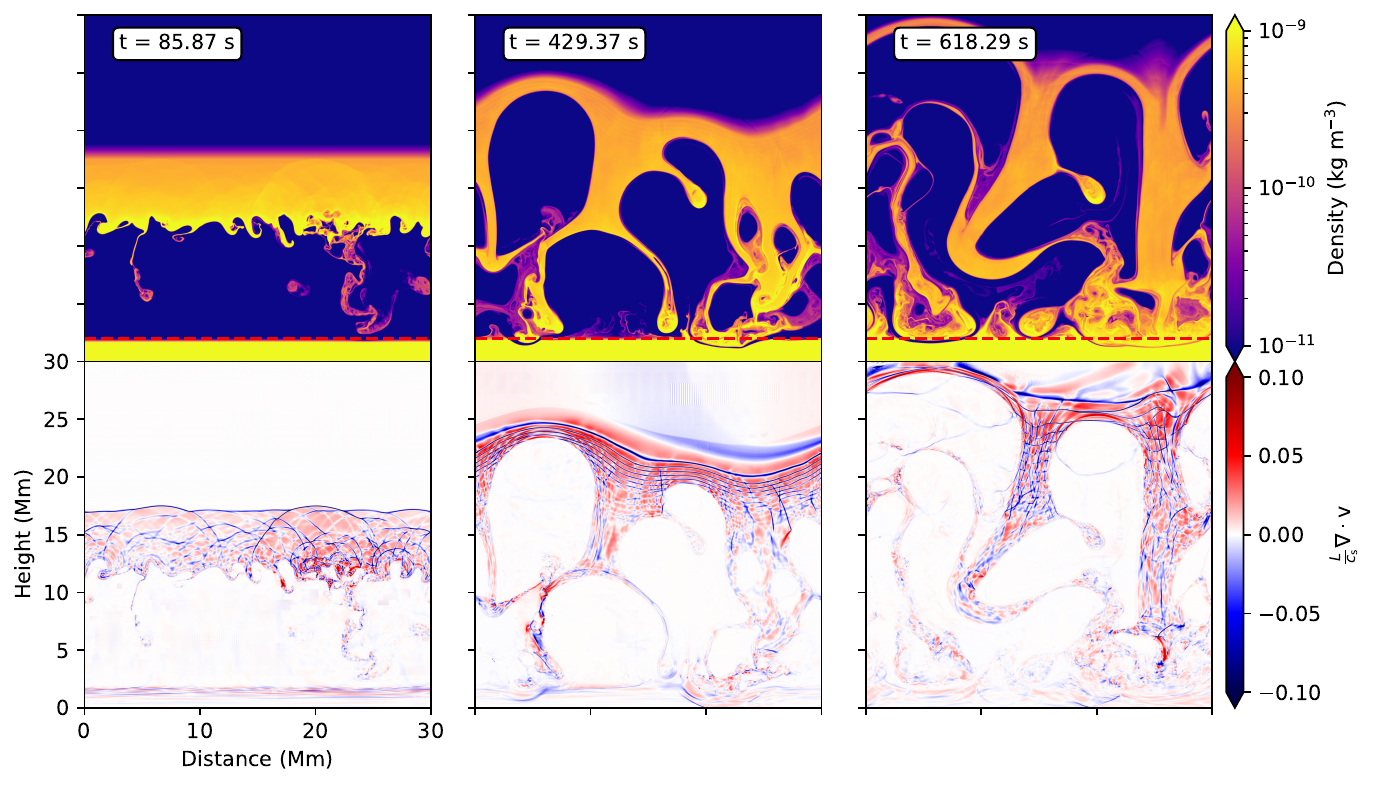}}

     \caption{Combined plots of $\log \rho$ and $\frac{L}{c_\mathrm{s}} \nabla \cdot \mathbf{v}$ at time 85.87~s, 429.37~s, and 618.29~s (left, middle, right) showing the evolution of bow shocks in the evolution of the solar prominence. The horizontal red dashed line indicates the 2~Mm threshold relevant for Figure~\ref{cooldense}.}
     \label{density_compressibility}
\end{figure*}

The equations (\ref{eqn:mhd1})-(\ref{eqn:mhd4}) are solved numerically on a dynamically changing grid structure, known as a hierarchical block-adaptive grid. The temporal integration strategy is a four$-$stage, third$-$order Runge-Kutta method, as described by \cite{ruuth2002two}. In the numerical flux computations, we employ the Harten-Lax-van-Leer approximate Riemann solver for multiple discontinuities (HLLD) following the approach proposed by \citep{miyoshi2005multi}. The HLLD method allows to accurately simulate high-resolution magnetohydrodynamic (MHD) evolutions while retaining positivity. Additionally, a third-order asymmetric slope limiter, originally introduced by \cite{koren1993robust}, is used in cell-center to cell-edge reconstruction. In our grid-adaptive simulations, refinement criteria are employed. These criteria are based on the instantaneous properties of the plasma and their gradients, following the methodology outlined by \citep{lohner1987}, and use the refinement criteria based on density. In the AMR code, the structure block-tree indicates that the option to refine or coarsen must be taken for each individual block \citep{2012JCoPh.231..718K}. In our simulation, we used a block size of $10 \times 10$. The generalized Lagrange multiplier (GLM) method was employed to control erroneous numerical magnetic field divergence, as described by \cite{2002JCoPh.175..645D}. A Courant value of 0.8 was selected in order to maintain temporal stability during the explicit time integration process. The initial setup and boundary conditions are identical to our previous 2D ideal MHD simulations from MC23. The introduction of uniform $\eta$ extends that previous setup to a resistive MHD simulation. The mean magnetic field strength is taken to be around 10~G. The initial magnetic field makes an angle ($\theta$) of  $\sim 88 \degree$ with the horizontal $x$-axis, so it is nearly perpendicular to the simulated plane. We use the same localized multi-mode perturbation as in MC23 to induce turbulence due to Rayleigh Taylor (RT) instability in the initial prominence setup and let the system evolve self-consistently. We note that without this perturbation, the prominence stays totally pressure-supported at its initial height. The boundary conditions are identical to our previous work, that is, periodic along the horizontal $x$-direction and a combination of hydrostatic and extrapolated in the vertical $y$-direction.

\section{Results}\label{s:results}
\subsection{Overall evolution of the prominence} \label{ss:evolution}

The evolution of density, as well as the corresponding compressibility quantification through the sound speed ($L/c_\mathrm{s}, L=\Delta_\mathrm{(x,y)}$) scaled divergence of the velocity field, for the resistive simulation is shown in Figure~\ref{density_compressibility} for time = 85.87, 429.37, and 618.29~s. From this figure, we can see how the quiescent prominence evolves from the onset of instability which occurs as a result of gravity's influence on an initial multi-mode perturbation and leads to the formation of vertical structures. The coherent structures, which manifest as pillars and bubbles, exhibit a progressive increase in fine-scale structures over time. Note that the presence and movement of (even very weak) shocks can be clearly identified in the velocity divergence. 

At time = 85.87~s, the deformation of the initial interface into vertical RT structures in the form of bubbles and pillars is in progress, with the nonlinear nature of the evolution already clear not only in the density distributions but also the presence of secondary compressive features within the prominence body, shown here in the bottom row of Figure~\ref{density_compressibility}. In particular, the $\nabla\cdot\mathbf{v}$ panel describes the formation of bow shocks above each of the rising plumes and falling fingers, visible as arcs.

From time = 429.37~s, we see that the cool and dense material of the prominence falls and collides the chromospheric boundary, such that it is partly reflected back to higher heights. Portions of the fallen prominence material remain in contact with the chromosphere but do not successfully mix. This deflection leads to the formation of more horizontal coherent structures within the prominence which was otherwise dominated by vertical structures in the longitudinal direction. At about this time (cf. MC23), the higher regions become beta unity and above, which results in the rising of the buoyant matter through our top boundary and fine-scale mixing, as the thermal pressure rises accordingly. In Figure~\ref{density_compressibility} (middle-right), shock waves which were seen as internal arch-shaped shocks in the initial stages, now appear to follow the large-scale deformations within the prominence material. The overall kink deformation of the prominence body, caused by the rising plumes, gradually displaces the upper prominence interface towards the upper boundary.

Finally, at time = 618.29~s, we see the complete development of the principle plumes and pillars of the nonlienar RT-unstable prominence. The falling fingers and rising bubbles are accompanied by a wealth of horizontal structures closer to the chromospheric boundary, filled with smaller scale structures that appear as discrete `cells' - not to be confused with the individual grid cells of the simulation. We will explore these features more in Section~\ref{sss:KHI}. An equal number of two large scale pillars and plumes can be discerned, with the latter now in the process of lifting a significant portion of the initial prominence mass through the upper boundary. In our previous study (MC23), we statistically explored the far nonlinear evolution of the RT-deformed prominence and its turbulent properties, but here we will focus only on this earlier phase of its evolution, where the prominence structure is still largely identifiable as seen in these figures.

\begin{figure}[]
        \centering
        
           \resizebox{0.8\hsize}{!}{\includegraphics{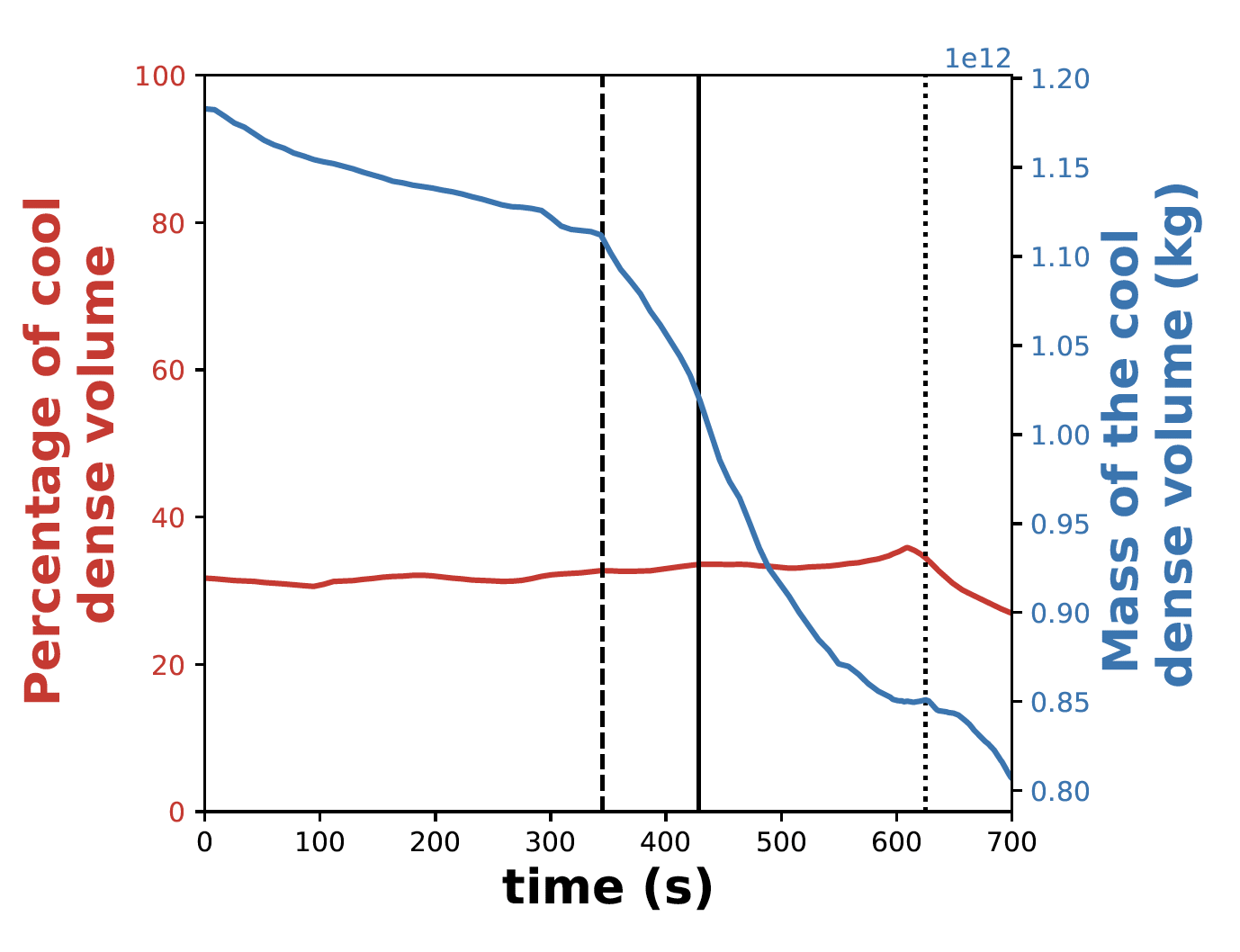}}
           
             \caption{Time evolution of the cool and dense solar prominence matter above the chromosphere boundary (above 2~Mm from the bottom boundary) taken below a threshold temperature of $5 \times 10^4$~K, and denser than $10^{-11}$~kg~m$^{-3}$. The dashed line represents the phase of the initial decrease of mass of the cool and dense volume in the solar prominence due to the merging of the cool dense material into the chromosphere boundary. The dotted line represents the phase where the prominence material rises towards the top boundary with a significant loss of the prominence material. The solid line represents the evolution at time = 429.37~s (see Figure~\ref{density_compressibility} (middle))}
             \label{cooldense}
\end{figure}

\begin{figure*}
        \centering        
           \includegraphics[width=0.9\textwidth]{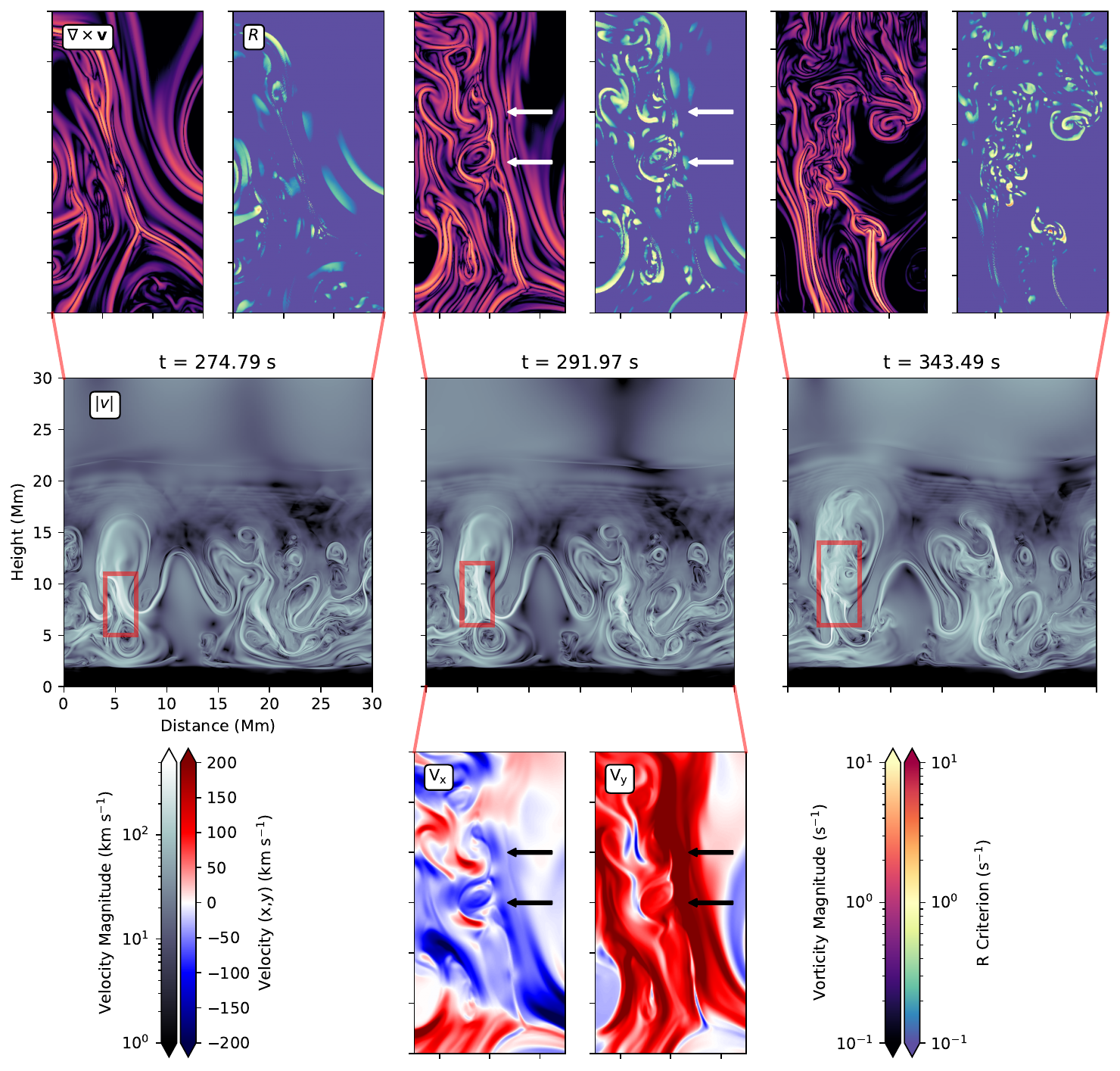}
           
             \caption{Evolution of secondary KH instability in the prominence at time = 274.79, 291.97, and 343.49~s as shown in the red insets. The top row shows the zoomed image for the KH instability regions, represented as both vorticity $\nabla \times \mathbf{v}$ and Rortex R, identified using the velocity magnitude in the middle row for each time. The bottom insets show the zoom region for time = 291.97~s, representing the $v_y$ and $v_x$ component of the velocity field at that time. Together, they signal a trail of (rising) vortex structures being formed.}
             \label{khi_ts}
\end{figure*}

Figure~\ref{cooldense} shows the evolution of the percentage of the cool dense prominence material within the simulation domain over time where the material below 2~Mm, the chromosphere, has been neglected. Cool and dense prominence material is defined as cells having a temperature below $5 \times 10^4$~K, and a density higher than $10^{-11}$~kg~m$^{-3}$. The dashed line in Figure~\ref{cooldense} represents the first strong instance of mixing of the prominence material around time = 350~s when two large pillars collide with the chromosphere boundary and result in the initial drop of the mass of the cool and dense volume. This decrease is entirely explained as the material breaching under the 2~Mm threshold. The solid line in Figure~\ref{cooldense}, represents the phase of the prominence at time = 429.37~s shown in Figure~\ref{density_compressibility} (middle). At this time, there is an ongoing exchange of material between the prominence and the chromosphere, with some being reflected back towards higher heights but the majority remaining below 2~Mm, hence the overall decrease. At the same time, the buoyancy of the rising bubbles is in the process of lifting the upper portions of the prominence towards the top boundary. The dotted line represents the prominence around time $\sim$ 620~s, when the initial horizontal prominence material is completely deformed due to the evolving shocks and rises towards the top boundary leading to an $\sim$ 28~\% decrease in the cool and dense volume of the prominence at that time. We have considered the evolution of the prominence due to the initial RT instability until time = 700~s, where the loss of the cool and dense volume is around $\sim$ 33~\% of the initial prominence material. As already discussed, we will therefore restrict our following analysis to this initial roughly 11-minute timeframe.

Throughout the initial evolution of our model, we find the formation of nonlinear structures from the initial stratification down to the small fine structures within the aforementioned `cells' of Figure~\ref{density_compressibility}. The cold and dense material is primarily located in the pillars that descend under the effect of gravity and hit the chromosphere-to-corona transition region. At this point, there is a clear change and accumulation of kinetic energy in the horizontal direction attributed to the non-linear stage development of the RT instability. Thereafter, vortices and swirling structures transfer energy among the scales and in both directions. In what follows, we will further explore the presence and characteristics of these complex plasma motions.

\subsection{Presence of secondary processes} \label{sss:KHI}
\subsubsection{The KH instability} \label{ssss:KHI}
\begin{figure*}
        \centering
       
           \includegraphics[width=0.9\textwidth]{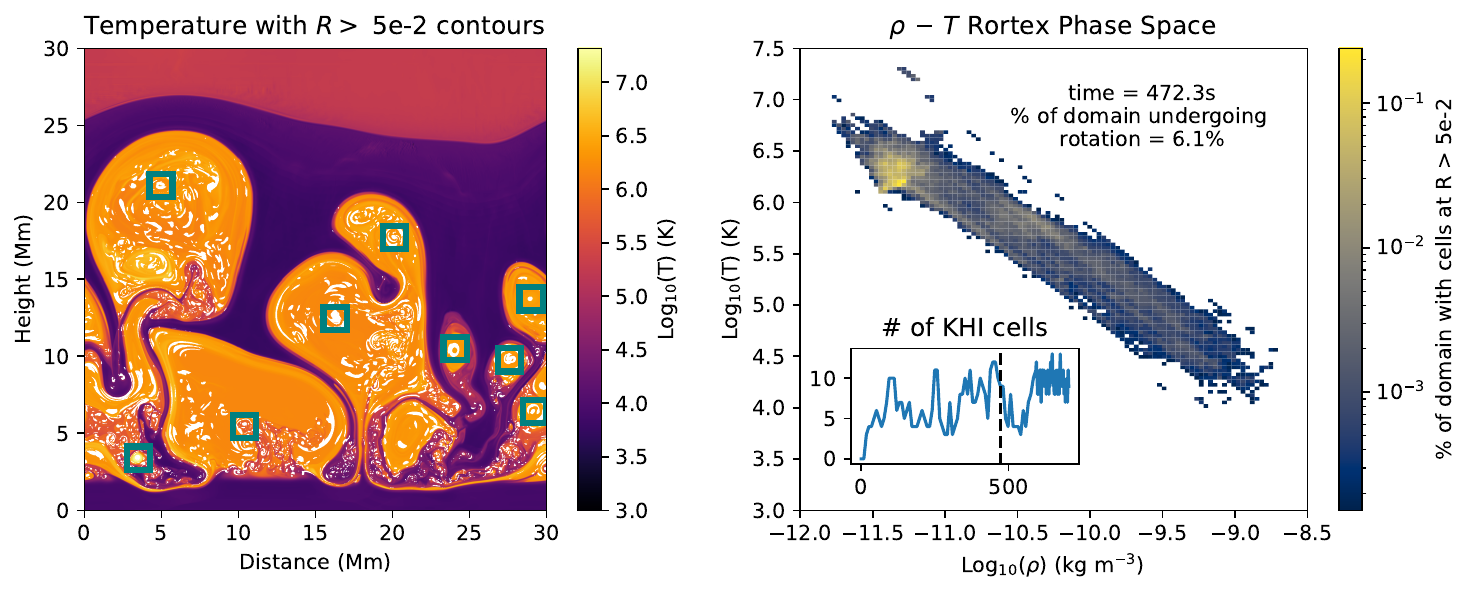}
           
             \caption{The condition of the plasma undergoing rotational motion within the domain, as highlighted according to the R-criterion. Left; the 2.5D domain of temperature with positions of $R>5\times10^{-2}$~s$^{-1}$ and where $\lambda_\mathrm{ci} > 10 \lambda_\mathrm{cr}$ overlaid in white, see text for description of these terms. A few clear KH instability cells have been indicated with teal boxes. Right; the density\,--\, temperature phase space of these extracted regions. A broad distribution is present, with a clear peak at the high temperatures and low densities of the solar corona. The inset axis plots the number of distinct KH cells in time (s), current time indicated by the dashed-black line, explanation in the text. An animation of this Figure will be available online.}
             \label{khi_stats}
\end{figure*}

From the initial time when the setup is in a slightly perturbed equilibrium state, the dynamics quickly become non-linear due to the multi-mode RT instability induced by the perturbation within the prominence-corona interface. In this early stage, we see the formation of falling fingers or pillars and rising bubble-shaped structures, in both cases there is obvious mixing at the interface layers. Under the effect of gravity, denser and cooler plasma precipitates towards the sun's surface, while buoyancy drives the less dense, hotter material upwards. This differential motion sets up convective patterns within the prominence and its coronal environment, thereby establishing shear flow layers.

Under hydrodynamic conditions, shear layers can be susceptible to KH instability if the velocity jump is sufficient. The differential motion of magnetised plasma induces flow-related instabilities, such as the KH type, that are sensitive to the magnetic field orientation; if $\mathbf{k}\cdot\mathbf{B} \neq 0$, where $\mathbf{k}$ is the wave vector, magnetic tension can suppress the development of rotational motion. In our simulation, the main magnetic field component is largely perpendicular to the simulated plane, and so we approximately recover the hydrodynamic behaviour in the sheared regions in the $(x,y)$ plane since $\mathbf{k}\cdot\mathbf{B} \approx 0$ and the restoring magnetic tension force is very weak. This is particular to our 2.5D setup, where invariance in the $z$-direction is adopted. 

Figure~\ref{khi_ts} shows the evolution of a localized shear flow interaction region that begins as an approximately laminar shear interface before fragmenting into discrete vortices. We show the velocity field magnitudes over the whole domain for time = 274.79, 291.97, and 343.49~s in its middle row. The top panels show a zoomed region manifesting KH instability, tracked using a combination of vorticity as $\mathbf{\omega} = \nabla \times \mathbf{v}$ on the left of each pair, and the so-called `R-criterion' on the right \citep[][sometimes referred to as a `Rortex']{canivetecuissa2022}. Here, following \citet{wang2019} and \citet{xu2019}, $R = |\omega| - \sqrt{|\omega|^2 - \lambda^2}$, with $\lambda = 2\lambda_\mathrm{ci}$, where $\lambda_\mathrm{ci} = \frac{1}{2}\sqrt{-[(u'_\mathrm{x}-v'_\mathrm{y}] - 4u'_\mathrm{y}v'_\mathrm{x}}$ for a negative discriminant and 0 otherwise, and $u'_\mathrm{x} = \frac{\delta u}{\delta x}$, etc. are the $\nabla \mathbf{u}$ 2D tensor components \citep[as in][]{canivetecuissa2020}. Although numerous formalisms exist in the literature to track the location, orientation, and strength of rotating fluids, such as $\mathbf{\omega}$, the Q-criterion, and $\lambda_\mathrm{ci}$, $R$ was constructed to be completely independent of the shear components that frequently complicate associated analysis \citep[][]{tian2018}. Locations of rotating fluid with very large radii of curvature, resembling sheared flows at the infinite limit, can then be excluded if the ratio of $\frac{\lambda_\mathrm{cr}}{\lambda_\mathrm{ci}}$, where $\lambda_\mathrm{cr}$ is the real solution to the associated eigenvalue analysis, is above some threshold, oftentimes~1. We set a threshold on this ratio to 0.1 to be even more restrictive. The bottom insets show the same zoomed region for time = 291.97~s, presenting the in-plane $v_x$ and $v_y$ component of the velocity field at that time. We find the vorticity to well describe the shear interface regions, with the additional R-criterion refining the spatial extent to just those features where rotation dominates over shear or general deformation. At this time, the utility of the R-criterion is clear as it refines the mapping onto portions of the discrete KH `rollups' identifiable by eye between 8 and 11~Mm. We note that local-box, two-fluid plasma-neutral simulations by \citet{Snow2024} recently highlighted the role of recombination and ionization processes in the thermodynamics of a KH shear layer. Future work could exploit this aspect in even higher resolution, global setups as realized here in single fluid settings.

In Figure~\ref{khi_stats}, we present a snapshot of the condition of the simulated plasma at time = 472.3~s in those locations identified by the R-criterion such that $R>5\times10^{-2}$~s$^{-1}$ where $\lambda_\mathrm{ci} > 10 \lambda_\mathrm{cr}$. In the left panel we see that many discrete locations satisfying the selection criteria are present across the simulation domain. In particular, at higher temperatures and this is reflected in the right panel presenting the 2D density\,--\,temperature histogram of those regions satisfying the selection criteria. The combination of higher temperature and lower density agrees with the 2D representation on the left, where it's clear the majority of the rotating plasma is located within the more coronal material, with some but far less present in the cooler prominence material itself. At this time, we see that 6.1\% of the 30$\times$30~Mm domain is undergoing rotational motion. In the animation that accompanies the Figure, we see that this fraction increases with time, peaking at 7.3\%, in agreement with the steady increase of secondary processes as the initially RT-unstable plasma develops mixing interface regions (cf. the \textit{prominence-corona-transition-region} (PCTR)) that in turn interact. In the left panel of Figure~\ref{khi_stats} and the associated movie, it is clear that smaller, discrete, and often short lived or `flickering' regions are identified by the R-criterion - in fact these features are present in all methods referred to above as a consequence of gradients on a discrete grid. Following authors such as \citet{yadav2020}, one may homogenise the signatures of rotating cells by smoothing the velocity field, and obtain a more robust estimate for the number of cells present within the simulation at a given time. The FWHM of the smoothing Gaussian is set assuming identification over four pixels at the IRIS platescale of 0.167$''$. We find the number of cells vary by a factor of two throughout the temporal evolution, peaking at $\sim$~10. Such a statistic is heavily influenced by the selection criteria, of course, where a broader smoothing to match the platescale of the \textit{Atmospheric Imaging Assembly} (AIA; \citealp{Lemen2012}) instrument onboard the \textit{Solar Dynamics Observatory} (SDO; \citealp{Pesnell2012}), for example, would lead to less identified cells.

The observational work of \cite{2018ApJ...864L..10H} shows the presence of KH instability vortices associated with downflows of the plasma in quiescent prominences. Different flow patterns were associated with the observed instabilities. Firstly, a coiling pattern as the blob falls and begins to wrap around the structure as well as a sinusoidal pattern that is mildly symmetric about the axis of the thread which moves upward. The velocity associated with the downflow pattern was observed to be 9\,--\,16~km~s$^{-1}$ whereas the velocity of the upflow sinusoidal pattern was 34~km~s$^{-1}$. The width of the first pattern was 0.9~Mm and the width of the second pattern was 0.48~Mm. The length covered by the thread throughout the process of rolling up in the downflow was 3.2~Mm and that of the sinusoidal structure was 2~Mm. At time 291.97~s, the zoomed insets in the bottom row of Figure~\ref{khi_ts} show the individual velocity ($v_x$ and $v_y$) components of our simulation. The $v_y$ component shows the vertically dominated shear flows while the $v_x$ component demonstrates the kind of coiling pattern found in observations. We find similar characteristics to \cite{2018ApJ...864L..10H} for those vortices evolving due to the KH instability, matching in particular the length scales with velocity of the same order of magnitude or higher. In the regions immediately surrounding the roll-ups, however, there are very high velocities of several hundreds of km~s$^{-1}$. The key difference here being that, and as generally summarised  in Figure~\ref{khi_stats}, the vortices present within our simulation are found in the much hotter $>10^5$~K plasma rather than the cooler prominence material. This will be explored more in Section~\ref{ss:synthetic}.

As the system evolves, this velocity shearing effect within the plasma may lead to the generation of high-energy events and structures such as plasmoids, current sheets, and jets. Additionally, KH instability may also foster the onset of tearing mode instabilities, a resistive MHD instability. This typically occurs on pronounced and thin two-dimensional (2D) current sheets, which in turn can be subject to secondary instabilities such as the aforementioned tearing mode. In the next section, we highlight how we can identify current sheets and get statistical info on their properties from our simulations. The interplay of these primary and secondary instabilities contributes to the complex dynamical evolution of the solar prominence environment. 

\subsubsection{Current sheets} \label{sss:current_sheets}

\begin{figure*}[]
        \centering
        
        \resizebox{1\hsize}{!}{ 
            \includegraphics[]{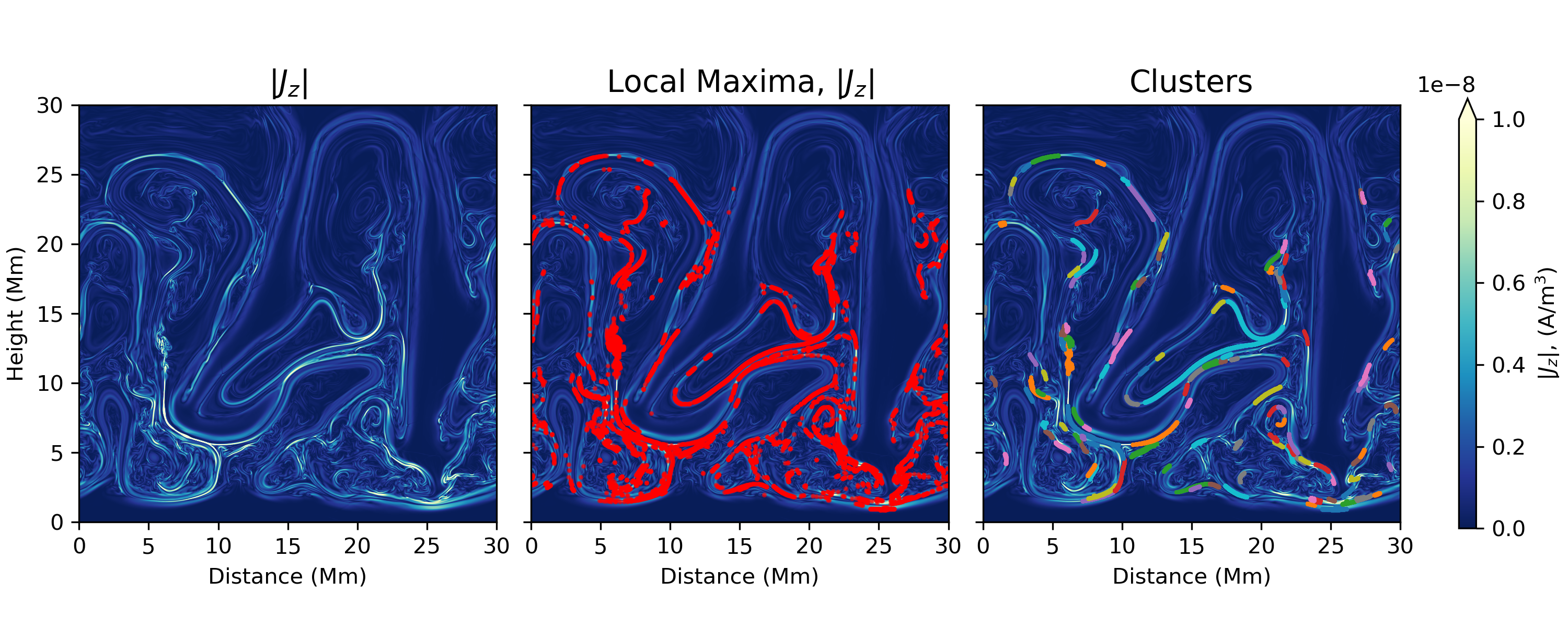}}
            
             \caption{Detection of current sheets using the modified algorithm based on current density $J_z$ at time 680.12~s. The left plot shows the current density magnitude, the middle plot shows the correspondence of local maxima events using a threshold higher than the mean $|J_z|$ with the current density ($|J_{z}|$), and the right plot shows the identification of clusters derived from density clustering methodology using local maxima events at time 680.12~s. 123 separate clusters could be identified.}
             \label{csheets}
\end{figure*}

The dynamic evolution of the solar prominence environment establishes the complex formation of current sheets, which are critical for magnetic reconnection and energy release. As we have discussed in Section~\ref{sss:KHI}, the preceding instabilities may play a significant role in the formation of these thin, high current-density regions through the deformation of magnetic field lines. The pathway towards prominence turbulence, driven by factors such as gravity, buoyancy, instabilities, and shear flows, creates a challenging backdrop for isolating the signatures of current sheets generated by these instabilities. The characterization of these instability induced current sheets is vital for understanding prominence stability, the triggering mechanisms of prominence eruptions, and the associated release of energy into the solar corona.

\paragraph{\textit{Methodology}}\mbox{}\\

The identification of current sheets in nonlinear plasmas is often nontrivial in full 3D, time-dependent settings. Here, we are in 2.5D, but still a rather nonlinear setting, and we are interested in locating fine-scale current structures that form due to evolving spatial eddies. These are formed due to the onset of KH instabilities which leads to a topological change of the magnetic fields. Identification of these current sheets is important to understand the energetic events associated with magnetic reconnection. \citet{2021A&A...651A..60H} showed the presence of current sheets and the observation of bi-directional jets in quiescent prominences. We will now evaluate whether we can use a clearly automatable procedure to identify and count current sheets in our snapshots.

\begin{figure*}[]
        \centering

            \resizebox{1.0\hsize}{!}{ 
            \includegraphics{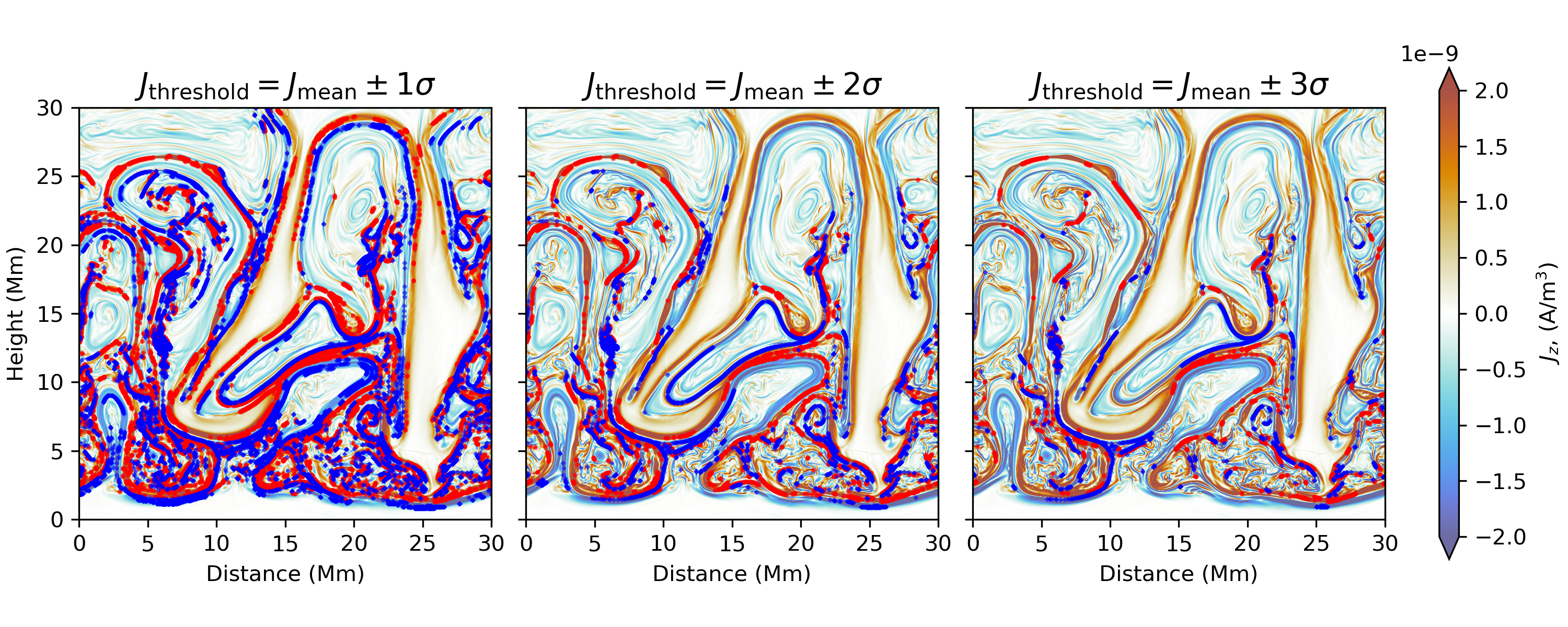}}

        \caption{Comparison of different threshold parameters of $J_{\textrm{threshold}}$ at time 680.12~s in correlating the local maxima events with the current density (|$J_z$|). The red current sheets represent the positive current sheets whereas the blue represent the negative ones. Depending on the threshold used, the algorithm may overlook significant current sheet regions to fulfill the parameter (right) or overfill the local maxima regions (left).}
        \label{comparison_sheets}
\end{figure*}

We follow the methodology given by \cite{2013ApJ...771..124Z} to identify local maxima associated with the current sheet formation in the MHD simulations, which suits our inhomogeneous, compressible, non-periodic boundary and gravity-driven evolution.  The initial procedure remains the same in determining the regions of local maxima in a designated snapshot, here we used time 680.12~s as a reference to first detect these current sheets. Figure~\ref{csheets} shows the methodology used to detect current sheets in the prominence. The current density for time 680.12~s is shown in Figure~\ref{csheets} (left). Given the distinctive features of current sheets, which are defined by extreme values in the current density profile, the task at hand is to identify local maxima in the magnitude of the current density. In order to accomplish this, the algorithm examines all data points that exceed a certain threshold current density, denoted as $J_{\textrm{threshold}}$. This threshold is set to be significantly higher than the average magnitude of current density across the whole dataset, we will discuss this later. The method then identifies the local maxima inside a cubic subarray around each of these selected locations. Each subarray is positioned at the candidate point with a radius of 3, a demonstration of the well-resolved peaks is shown in the middle panel of Figure~\ref{csheets} denoted by the red blobs. Each maximum is thereafter associated with a current sheet denoted by a current sheet index $i$, and the related current density is referred to as the peak current density, $J_{\textrm{max},i}$. The right panel of Figure~\ref{csheets} shows the correlation of the local maxima events at time 680.12~s, to the magnitude current density of the simulation at that time. We will now need to group these points into individual current sheets, requiring us to cluster nearby points and then derive the sheet typical dimensions (width and length).

\begin{figure}[htbp!]
        \centering

        \resizebox{1.0\hsize}{!}{ 
            \includegraphics{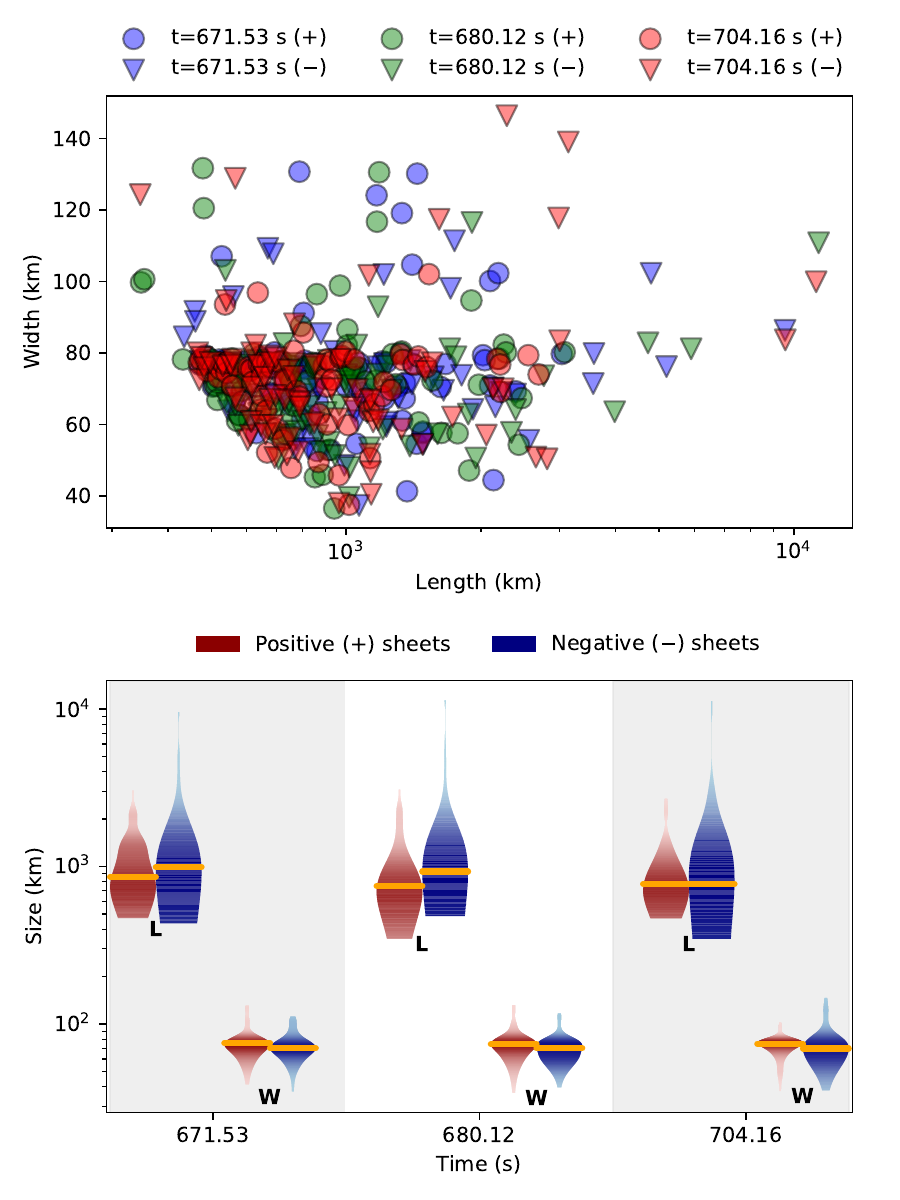}}

        \caption{Top: Scatter plot of all current sheets characterized in length being longest dimension (horizontal) versus width (vertical) for three times: 73 clusters (positive) and 65 clusters (negative) for t= 671.53~s, 75 clusters (positive), and 62 clusters (negative) for t = 680.12 s, and 73 clusters positive and 70 clusters (negative) for t = 704.16~s (blue, green, red, respectively). The `+' refers to the positive current sheets while `-' refers to the negative current sheets. Most of the current sheets are smaller ranging between 631~--~1388~km in length and of 61~--~78~km in width. The largest current sheet analyzed by the algorithm in the simulation goes up to $\sim$~11355~km in length and around $\sim$~116~km in width.\\
        Bottom: Statistical distribution of current sheet lengths and widths at the same times, shown as violin plots for positive (red) and negative (blue) current sheets. The width of each violin indicates the probability density estimated by kernel density estimation, while the solid orange horizontal lines mark the medians. The “L” and “W” labels denote distributions for length and width, respectively. Shading is used to visually separate the three time instances. These distributions reveal asymmetries between positive and negative sheets in both length and width, highlighting temporal evolution and polarity dependence in current sheet morphology of the simulation.}
        \label{scatter_sheets}
\end{figure}

\begin{figure*}
        \centering
        \FIG{
            \resizebox{0.9\hsize}{!}{ 
            \includegraphics[]{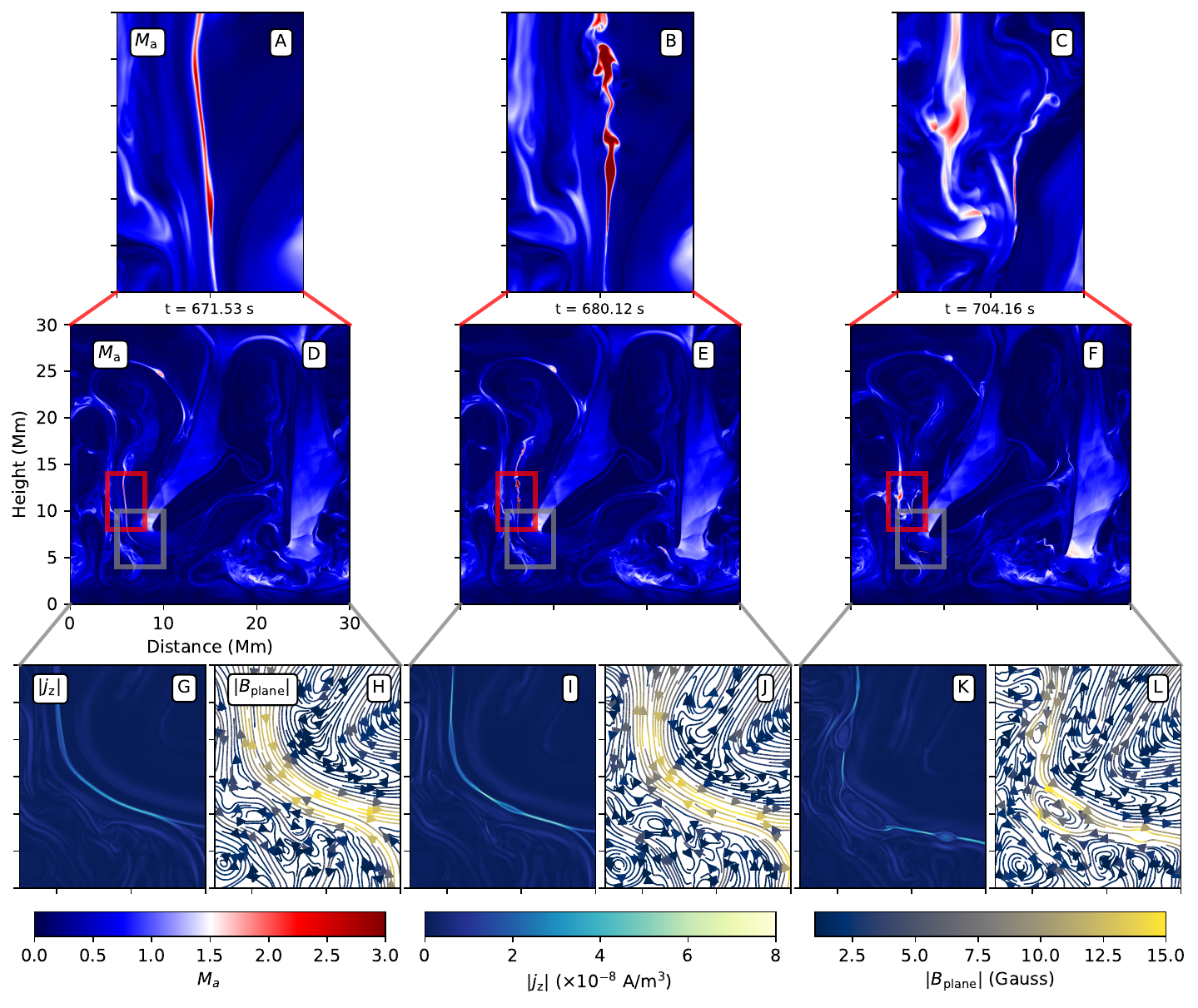}}
            
            }
             \caption{Time series of the occurrence of jets in the prominence identified by regions where $M_A > 1$ at time = 671.53, 680.12, and 704.16~s. Panels (A\,--\,C) show the zoomed images of Panels (D\,--\,F) and the occurrence of jets, here essentially flowing upwards. Panels (G\,--\,L) show the time evolution of current density ($|J_z|$) and magnetic field ($|B_\mathrm{plane}|$) for zoomed regions where the jets emerge. Both the current sheets and subsequent magnetic islands responsible for the jets are clearly visible.}
             \label{jets}
\end{figure*}

The next step is to find clusters around the local maxima points iteratively and then identify the points belonging to each cluster that lie above a certain threshold. We augmented the previous identification algorithm to use a clustering technique called density-based spatial clustering of applications with noise \citep[DBSCAN;][]{10.5555/3001460.3001507} to find the clusters representing the current sheets. The DBSCAN method is a density-based approach that identifies and groups data points inside a specified space based on their proximity to one another. It forms aggregates by collecting points that are closely clustered together. The DBSCAN algorithm utilizes a distance function to determine the proximity of data points. In this implementation, a neighborhood growth radius of four grid points is used to expand regions around points of interest. The DBSCAN parameter $\epsilon$, set to three grid points, defines the clustering neighborhood size, while the minimum number of points is set to one to permit small spatial jumps and distinguish a cluster from isolated outliers. We then group those points in the neighborhood that are half of the maximum current density threshold of that given cluster, $J_{\textrm{min},i}=J_{\textrm{max},i}/2$, and further exclude point identifications and very short current sheets with an area below 3.3~Mm$^2$. We find 123 individual clusters which are the current sheets for the reference time as shown in Figure~\ref{csheets}. The choice of a threshold parameter and number of neighborhood cells can greatly affect the detection of individual current sheets. Using the parameters above, a visual comparison demonstrates that this approach accurately labels most of the extended $|J_\mathrm{z}|$ ridges as current sheets with few exceptions \citep[cf.][]{Lapenta2022}.

The current sheets in the simulations are evolving structures that show varying current density intensities in an overall dynamic environment. To determine the properties of these evolving current sheets, we consider $J_{\textrm{threshold}} = J_{\textrm{mean}} + n \sigma$ rather than a static value. In doing so, we maintain robustness and consistency in detecting the current sheets in time. We experimented with different settings for this parameter, a summary of which is shown in Figure~\ref{comparison_sheets}. A choice of $n=1$ shows a tendency to overfit in the sense that many local, often singular maxima are identified within the highly dynamic, evacuated regions in between descending fingers, while a value of $n=3$ neglects many of the more elongated current sheets that drape around the fingers and plumes. For our study, we decided to work with the intermediate parameter value of $n=2$, a tradeoff which captures both elongated and localized current maxima satisfactorily, and importantly does so throughout the simulation run (not only the time shown in Figure~\ref{comparison_sheets} and evidenced in Figure~\ref{csheets}). Hereafter, we make a distinction between positive and negative current sheets, as in Figure~\ref{comparison_sheets} (red and blue), to find the individual clusters and avoid biasing caused by $|J_\mathrm{z}|$ grouping closely co-located but oppositely oriented sheets as single sheets. Of course, the bias of choosing $n=2$ remains.

To characterise the evolution of the current sheets quantitatively, we follow \cite{2013ApJ...771..124Z}. The largest distance between any two pairs of points in the current sheet's $x-y$ cross-section, given by the maximum of the euclidean pairwise distance map, is how we define the length. We then use a method similar to \cite{2010PhRvE..82e6326U} to estimate the second dimension (width). By assuming that the current sheet is mostly uniform along its extent, i.e., no taper, and given our definition of a current sheet considers those adjacent cells for which $J_{\textrm{min},i}=J_{\textrm{max},i}/2$ is satisfied, dividing the cross-sectional area of a given cluster's concave hull by its length yields an average width according to the full-width at half-maximum (FWHM) of the cluster \citep[][]{Barber2013}. 

\paragraph{\textit{Statistics}}\mbox{}\\

We characterised all extracted current sheets present at times t= 670.17, 680.12, and 685.60 s, shown in Figure~\ref{scatter_sheets} (top) as a scatter plot of width (the sheet shortest dimension) versus length (the sheet longest dimension). The positive current sheets are marked by `+' and the negative ones by `-'. We found 73 and 65 clusters for t= 671.53~s, 75 and 62 clusters for t = 680.12~s, and 73 and 70 for t = 704.16~s for positive and negative sheets, respectively. Note that the distinction of positive and negative sheets understandably leads to a higher number of extracted current sheets than $|J_\mathrm{z}|$ in Figure~\ref{scatter_sheets}. The majority of the clusters found by the algorithm have lengths ranging between between 631~--~1388~km in length and 90~--~152~km in width. The smallest current sheets appear to cluster in tight distributions within the simulation domain, and likely contribute more heavily to the dissipation of stressed magnetic energy (left panel of Figure~\ref{comparison_sheets}. The largest current sheet found goes up to 11355~km in length and around 116~km in width.

The violin plots in Figure~\ref{scatter_sheets} (bottom) display the distributions of current sheet lengths (L) and widths (W) separated by polarity and time in the simulation. Each violin visualizes a kernel density estimate of the probability density function of current sheet sizes.

At t=671.53~s, positive polarity length distributions peak near a median of $\sim$~861.8~km with an interquartile range (IQR) from $\sim$~692.3~--~1352.3~km, while negative sheets exhibit slightly larger typical lengths with median $\sim$~991.4~km (IQR 663.3~--~1544.8~km). However, the corresponding sheet widths are tightly clustered: for positive sheets, the median is $\sim$75.6~km (IQR: 67.2~--~78.0~km), and for negative sheets, $\sim$~70.2 km (IQR: 65.5~--~77.5~km). The width distributions are notably narrow compared to the broader length distributions, indicating far less spread and the absence of extreme outliers, which reflects the asymmetrical temporal evolution of the current sheets.

At t=680.12~s, the distributions shift toward slightly smaller and more uniform scales. The length medians shift slightly downward: positive sheets show a median length of $\sim$~752.3~km (IQR 573.7~--~1134.5~km), while negative lengths hold a median of $\sim$~930.8~km (IQR 708.6~--~1228.0~km). Width remains tightly distributed with medians of $\sim$~74.4~km (IQR: 65.5~--~78.6~km) for positive, and 70.2~km (IQR: 60.2~--~76.3~km) for negative sheets reflecting increased variability in sheet width. This temporal evolution suggests a gradual shortening and narrowing of positive sheets, while negative sheets remain comparably larger in size.

By t=704.16~s, the typical sizes converge yet further: positive and negative sheet lengths have nearly equal medians ($\sim$~775.7 and $\sim$~774.3~km, respectively) and overlapping IQRs. Width distributions also narrow and align closely around $\sim$~74.9~km (IQR: 67.9~--~77.4~km) for positive and $\sim$~70.0~km (IQR: 59.9~--~78.5~km) for negative sheets. This convergence likely reflects the dynamic balance reached in sheet formation and dissipation as the turbulent phase develops.

The shapes and widths of the violins additionally reflect the length distributions being broader and more skewed compared to width distributions which are more sharply peaked, as seen in the top pane of the same Figure. Lengths exhibit a tail towards larger values, consistent with elongated resistive layers, whereas widths remain relatively constrained and reflect typical properties governed by the chosen $\eta$.

Overall, these violin plots quantitatively establish the temporal evolution and polarity-dependent morphology of current sheets in the simulation, highlighting weak initially marked asymmetries - likely inherent to the specific initial condition and nonlinear development i.e., clockwise vs counterclockwise KH instability development - that approach statistical similarity as time progresses and the imprint of the linear\,--\,nonlinear development is lost. Such a mild trend should be explored at higher resolutions and at corresponding $\eta$.

\subsubsection{Jets} \label{sss:jets}
One of the energetic events associated with the occurrence of current sheets that demonstrate magnetic reconnection are jets. Observational work of \cite{2021A&A...651A..60H}, has revealed the presence of bi-directional jets in solar prominences using IRIS slit-jaw imagers. In the current study, the shearing motion of plasma due to the KH instability is one of the main reasons for forming current sheets and associated jets. In Figure~\ref{jets}~(D)-(F), we show the time evolution of jets which we identify based on the Alfv\'enic Mach number 
\begin{equation}
    M_A = \frac{|v|}{v_a}\,,
\end{equation}
where $v$ is the flow speed of the plasma and $v_a = \frac{B}{\sqrt{\mu\rho}}$ is the Alfv\'en velocity. We look at in-plane velocity, while $B$ is the total magnetic field strength and $\rho$ is the density of the plasma. If the value of $M_A > 1$, the flow is super-Alfv\'enic. Super-Alfv\'enic outflows are usually a clear indication of local reconnection events. These jets are found to be short-lived events in our simulations (Figure~\ref{jets}~(A\,--\,C)) having a lifetime of around 18~s. They show sideways motion as they rise up from the points where current sheets deform due to tearing-like events at magnetic-reconnection sites in the simulations. Figure~\ref{jets}~(G\,--\,L) shows that the jets originate from pronounced current sheet regions as quantified by the current density field ($|J_z|$). The evolution of these current sheets gives rise to reconnection regions that lead to the jet exhausts shown in Panels (A\,--\,C). The reconnection is essentially component-reconnection, noting that the main magnetic field component $B_z$ stays unidirectional out of the plane, but is clearly rearranging in-plane magnetic field ($B_x,B_y$). In the examples shown here, the jets mainly occur near a pronounced, curved current sheet located at about $x,y = (6,7)$~Mm. For the duration that we have studied, we identify two main jet exhausts that follow in quick succession over a period of around one minute, each comprised of many components in the fine structure as the initial laminar flow becomes unstable.

\subsection{Synthetic observations} \label{ss:synthetic}
Prominences are frequently observed in extreme ultraviolet (EUV) using the EUV instruments from the space-based SDO/AIA, and also in the hydrogen H$\alpha$ spectral line using optical instruments from ground-based observatories. To compare the spatio-temporal dynamics found in our simulations with the motivating observations of e.g., \citet{hillier2021jets}, we have made synthetic observations of our simulations. In order to do so, the radiative transfer equation for the relevant elements and ionizations is required. From \citet{1986rpa..book.....R}, the general solution to the radiative transfer equation is written as:
\begin{equation}
    I_{\lambda} = I_{\lambda}(0) e^{-\tau_{\lambda}} + \int_0^{\tau_{\lambda}} e^{-(\tau_\lambda - \tau_{\lambda}^\prime)} S_\lambda(\tau_\lambda^\prime)d\tau_\lambda^\prime,
    \label{eqn:source_function}
\end{equation}
where $I_\lambda(\tau_\lambda)$ is the intensity of measured light, of wavelength $\lambda$, $\tau_\lambda$ is the total optical thickness along the chosen line of sight (LOS), and $I_\lambda(0)$ is the intensity of any background illumination. The source function is defined as $S_\lambda = \frac{j_\lambda}{\alpha_\lambda}$, where $j_\lambda$ and $\alpha_\lambda$ are the emission and absorption coefficients, respectively, dependent nonlinearly on temperature and density. They are the measure of the addition/removal of intensity to/from the light due to the local plasma conditions having local optical thickness $\tau_\lambda^\prime$. The implementation here represents a similar approach to that of \citet{Jenkins:2022}.

\begin{figure}[]
        \centering
        \FIG{
           \resizebox{\hsize}{!}{ 
           \includegraphics[clip=,trim=30 30 20 0]{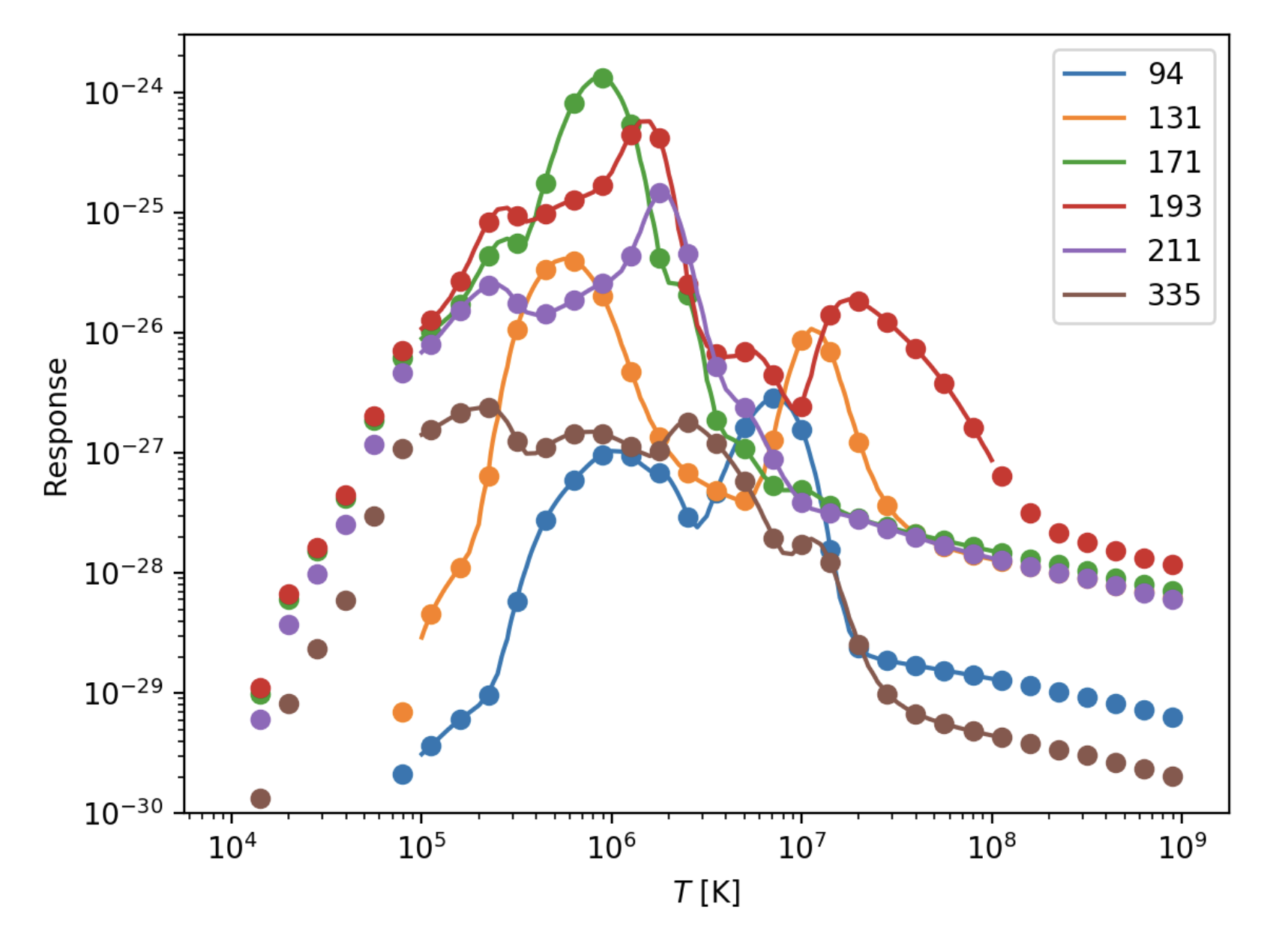}}
           }
             \caption{The temperature response function of the different AIA passband filters as in \citet{will_barnes_2022_6640421}.}
             \label{AIA_response_function}
\end{figure}

\begin{figure*}[]
        \centering
            \resizebox{0.95\hsize}{!}{ 
            \includegraphics[]{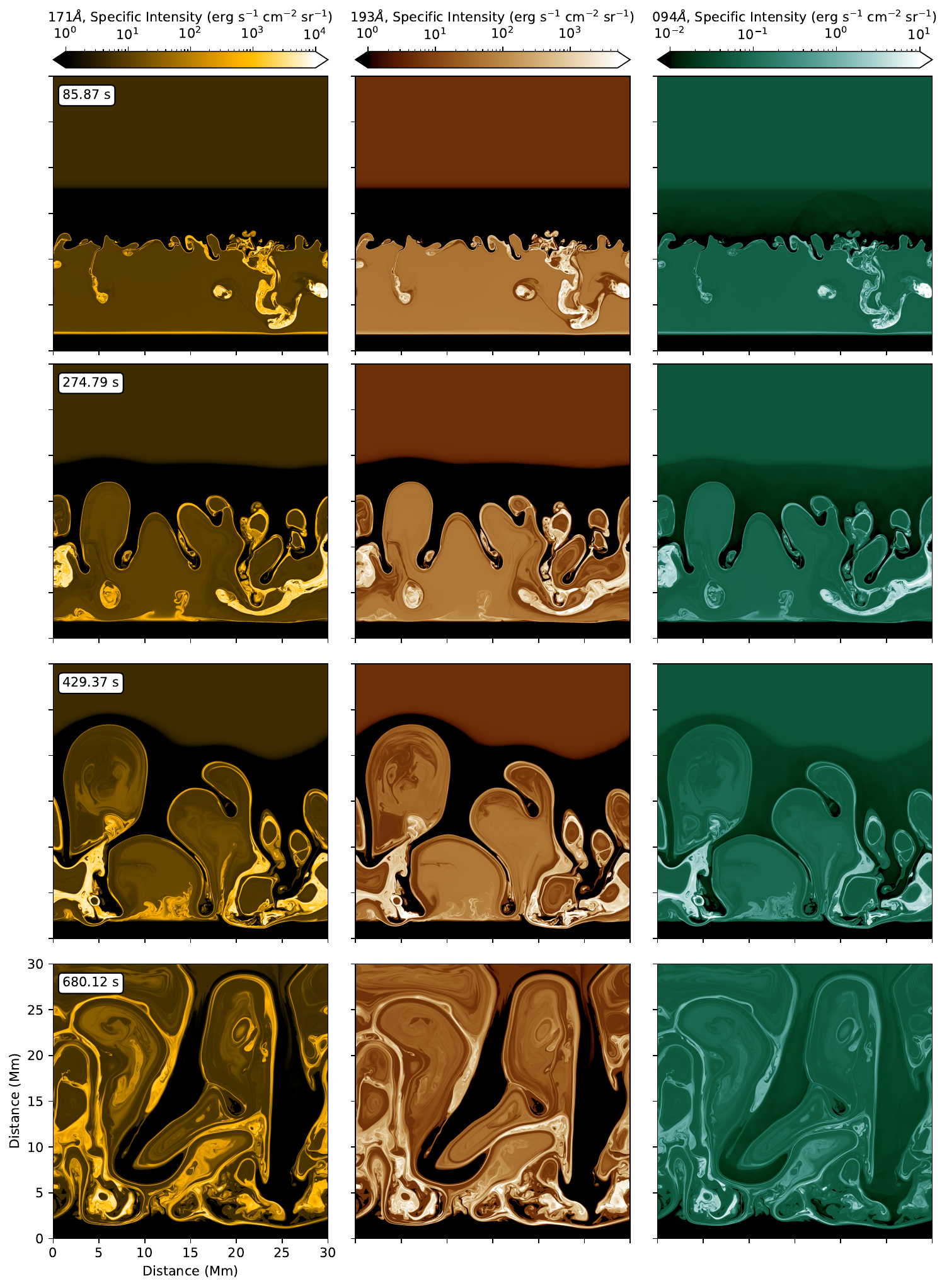}
            }

             \caption{Specific intensity counterparts to the simulation, where we account for emission and absorption. Synthetic images for the broadband 171 (left), 193 (middle), and 094~\AA\ (right) SDO/AIA filters are shown for time = 85.87, 274.79, 429.37, and 680.12~s from top to bottom respectively. }
             \label{synthetic_images}
\end{figure*}

\begin{figure*}[]
        \centering
        \resizebox{1.0\hsize}{!}{ 
            \includegraphics[width=1\textwidth]{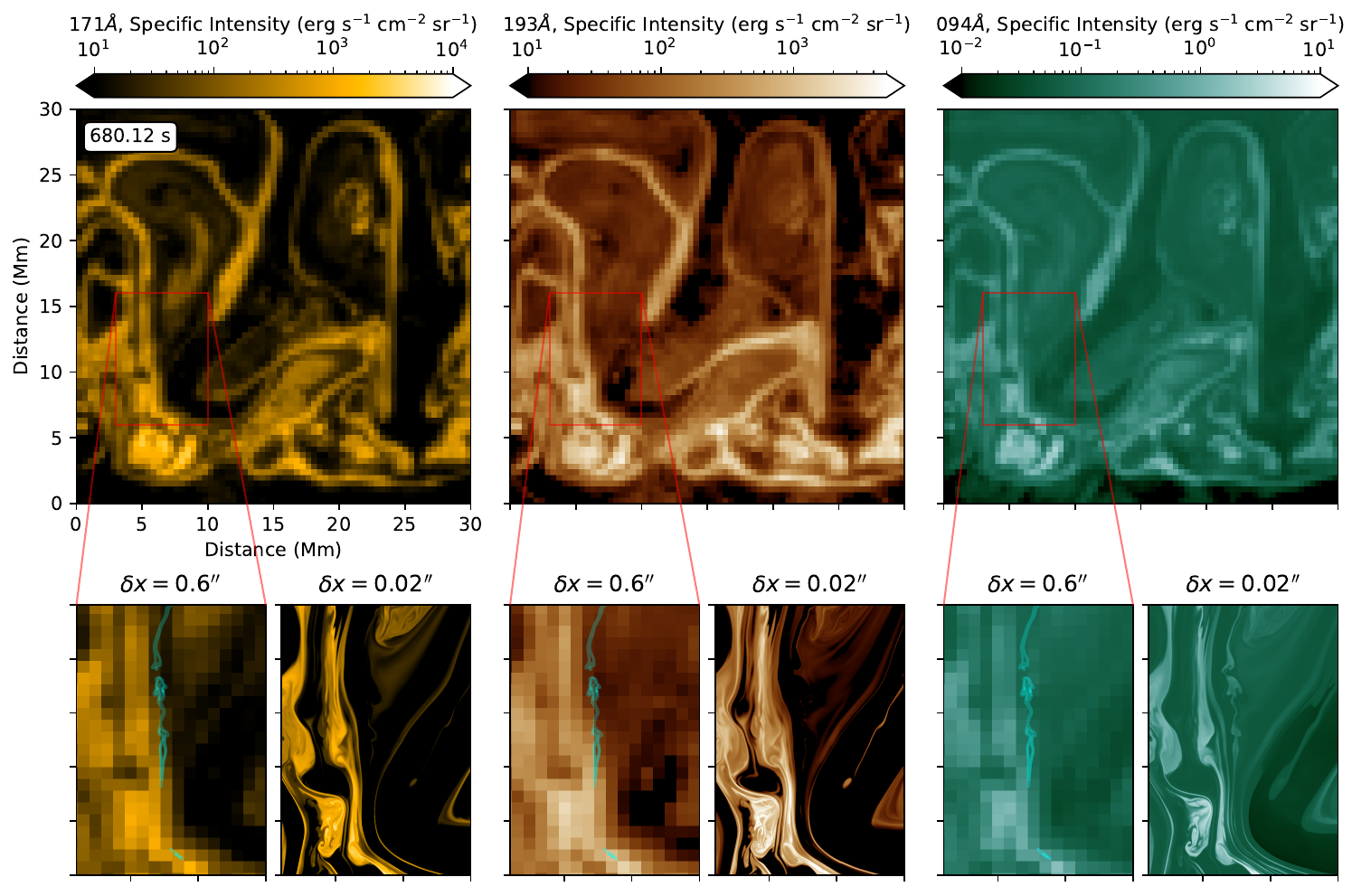}}

             \caption{Synthetic intensity counterparts to the simulation for AIA 171, 193, and 094~\AA
 at 680.12~s, convolved and de-resolved with the 0.6$''$ platescale PSF of the AIA instrument on board SDO. Lower row zooms in on the jet indicated by a high $M_A$ in Figure~\ref{jets}~(D)-(F) (overlaid and contoured in cyan); the left and right panels of each passband are at the AIA 0.6$''$ and \texttt{AMRVAC} 0.02$''$ resolution, respectively.}
             \label{AIAres_171}
\end{figure*} 

To qualitatively compare this work against observations, we synthesize images in three of the passbands observed by AIA, namely 171, 193, and 094 \AA. These passbands were chosen as the simulation contains temperatures between 10$^3$\,--\,10$^7$K and their temperature response functions cover a range peaking at $\approx$ 0.8, 1.5, and 7 MK, respectively, as shown in Figure~\ref{AIA_response_function}. Plasma emitting around these temperatures will contribute intensity to the synthetic observation accordingly. The physical variables of plasma density and temperature were converted into these spectroscopic observables under a combination of the `coronal approximation' and local thermodynamic equilibrium (LTE) \citep[][]{Verner:1996,Keady:2000}. Generally, coronal emission will appear bright in the resulting images, whereas prominence material will appear dark. Additionally, photoionisation of hydrogen and helium by such emission ($j_\lambda$) is responsible for active absorption ($\alpha_\lambda$) \citep[][]{Kucera:2015}. As these simulations are in 2.5D and hence contain no background, we assumed a geometrical integration depth of 500~km, and a background illumination of 1~$\times$~10$^{1}$, 0.5~$\times$~10$^{1}$, and 5~$\times$~10$^{-2}$~erg~s$^{-1}$~cm$^{-2}$~sr$^{-1}$ for 171, 193, and 094 \AA, respectively. Figure~\ref{synthetic_images} then shows the synthesised prominence appearance for times = 85.87, 274.79, 429.37, and 680.12~s for the three aforementioned AIA passbands.

From time = 85.87~s, the early development of the simulation describes two primary domains of darker (cooler) and brighter (mixed) material. In all passbands, the clear absorption of the prominence body captures the early nonlinear RT instability with dark falling fingers developing along the previously plane-parallel interface. The emitting features are that of prominence material that has already mixed with plasma initially at coronal temperatures, leading to dense plasma at hotter temperatures than in the prominence interior appearing bright in each of the passbands \citep[cf.][and the PCTR phenomenon]{Hillier:2023}. By time = 274.79~s, rotating blobs that have separated from the cool and dense prominence layer create bright shear interfaces throughout the region between the prominence and the chromosphere. Then, at time = 429.37~s, it is well-captured how the falling prominence material interacts with the underlying chromosphere, with a significant portion being reflected back upwards and sidewards at the same time as inducing additional rotation. At this time, the plasma mixing caused by the combined nonlinear developments of the RT and KH instabilities leads to the development of additional bright fine structures for each of the passbands. By time = 680.12~s, the dual action of the KH instability and magnetic reconnection has isolated structures into identifiable `cells', which can be contrasted to density views as seen in  Figure~\ref{density_compressibility}.

To facilitate a qualitative comparison against observations, we have convolved the synthesis with the 0.6$''$ platescale PSF of each of the AIA passbands \citep[][]{Grigis:2012,Barnes:2020}. The result, shown in Figure~\ref{AIAres_171}, significantly complicates any further interpretation, in particular for the jet-like feature of Figure~\ref{jets}. The bottom-right panel of Figure~\ref{AIAres_171} compares the synthesis at both the \texttt{AMRVAC} and AIA resolutions for the 094$\AA$ passband, with contours of $M_A>1$ overlaid in cyan. At the \texttt{AMRVAC} resolution, the jet is visible in the 094~\AA\ synthesis whereas at AIA resolutions all trace of it has been lost. The jet is entirely not visible in the 171 or 193~\AA\ panels of the same figure, at either \texttt{AMRVAC} or AIA resolutions, as the thermodynamics of the jet are incompatible with the formation temperatures of these passbands. In all panels, the hotter material that bounds the individual rotating `cells' remains visible, clearly demarcating distinctly different regions within the nonlinearly evolving RT and KH instabilities, in particular the largest of such examples.

Ground-based observatories commonly observe, amongst others, the strong Hydrogen $n = 3 \rightarrow n=2$ (H$\alpha$) line at 6563 \AA. To synthesize the H$\alpha$ emission and absorption we follow \citet{2015A&A...579A..16H}, who presented a method to convert the local pressure and temperature values within the plasma to H$\alpha$ opacity ($\alpha_\lambda$), a relationship obtained from a comprehensive array of 1.5D radiative transfer models. Whilst an approximation, we have already demonstrated its accuracy in comparison to non-local thermodynamic equilibrium (NLTE) synthesis applied to more complex models \citep[][see also \citealp{Osborne:2024}]{Jenkins:2023}. A core consideration within these models is the assumption of a constant source function of Eq.~\ref{eqn:source_function} along a given LOS, yielding a simplified equation,
\begin{equation}
    I_\lambda = I_\lambda(0)e^{-\tau_\lambda} + S_\lambda (1-e^{-\tau_\lambda})\,.
\end{equation}
The resulting emergent (specific) intensity of the H$\alpha$ line is therefore found through a LOS integration of the approximate $\alpha_\lambda$ according to the tables of \citet{2015A&A...579A..16H}, wherein a coarse height-dependent estimation to the value of $S_\lambda$ is also provided.

\begin{figure*}[htbp!]
        \centering

            \resizebox{1.0\hsize}{!}{ 
            \includegraphics[width=1\textwidth]{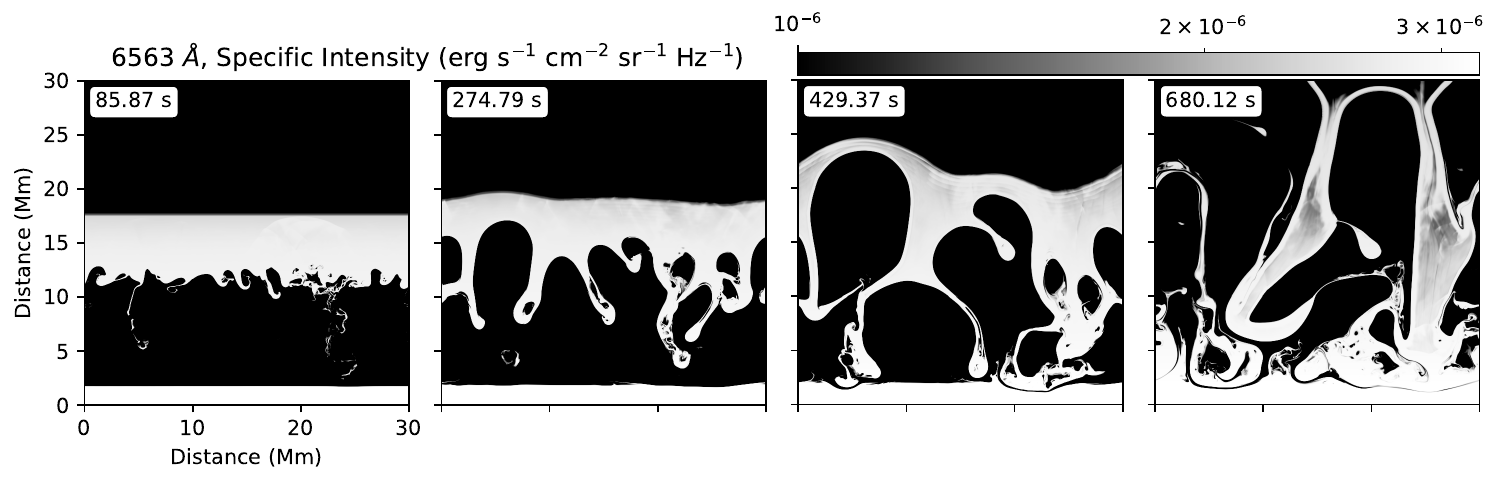}}

             \caption{Synthetic images for the narrowband Hydrogen-H$\alpha$ filter, for a LOS view into the plane of the simulation at time = 85.87, 274.79, 429.37, and 680.12~s.}
             \label{halpha_images}
\end{figure*}

As already discussed, the dark features that are observed in the EUV passband images are due to the photoionisation of the cooler Hydrogen and Helium material, and these are the prominence falling pillars. This cool plasma with temperatures $\sim 10^4$~K instead appears bright in the equivalent H$\alpha$ images. Figure~\ref{halpha_images}, shows the evolution of our prominence according to H$\alpha$ synthesized images for times 85.87, 274.79, 429.37, and 680.12~s. To first order, the locations of the dark features within the EUV synthesis are cospatial with the brighter structures in H$\alpha$. In particular for the snapshot at time = 680.12~s, however, the relationship between the EUV and H$\alpha$ features is not so trivial, with overlap found for even some of the brighter EUV pixels. Indeed, depending on the moment within the simulation, some of the material rotating due to shear flows is simultaneously visible in both the EUV and H$\alpha$ syntheses. As already discussed, however, the hot jet feature is not present within the H$\alpha$ panel at time = 680.12~s. Interestingly, the shocks of Figure~\ref{density_compressibility} are present throughout the prominence, evidenced by the formation of density compression layers that are subsequently encoded within the H$\alpha$ emission due to its sensitivity to the local plasma pressure quantity \citep[][]{2015A&A...579A..16H}. Depending on the applicability of the reduced H$\alpha$ synthesis approach, and the 2D versus 3D simulation setup, similar features may be visible within observations \citep[][]{Jenkins:2023}.

\section{Discussion} \label{s:discussion}

    In this manuscript, we have performed a resistive extension to the 2.5D simulation of MC23 and detailed the subsequent evolution considering both theory and observations. MC23 advocated for an immediate 3D extension to the prominence model, but we here maintain the 2.5D geometry so as to study the resistive behaviour in isolation from additional out-of-plane dynamics. The prominence here evolves from the same initial multimodal perturbation triggering the RT instability and leading to the development of distinct `falling fingers' and `rising plumes'. The subsequent interaction of the falling fingers with the underlying chromosphere strongly influences the overall evolution, generating a significant horizontal momentum and forcing the collision of adjacent structures \citep[in contrast to][]{Rees-Crockford:2024}. Secondary processes are then formed, with both the KH instability and magnetic component reconnection driving additional dynamics. For an identical initial condition and numerical scheme, the markedly different distributions of plasma when compared to MC23 (see their Figure~3) is attributed to the inclusion of finite resistivity throughout the domain. For an equivalent timestep, the model presented here yields more coherent structures, whether in `fingers' or `plumes'. This distinction is a function of the magnitude of $\eta$ chosen; a necessarily large value of $10^{-4}$ ($\gg10^{-9}$) was set so as to resolve its influence on the discrete grid, consequently introducing a minor diffusive smoothing of the thermodynamics that becomes more noticeable over longer timescales \citep[cf.][]{Ripperda:2019b}. We note that compared to MC23, which only had numerical resistivity at play, we here targeted actually resolved resistive evolutions (at unrealistically low magnetic Reynolds number), using simulations with a higher effective resolution.

    In quiescent prominences, the primary RT instability initiates upflows and downflows, successfully reproduced here in our simulation as shown in Figure~\ref{cooldense} \citep[][]{hillier2012numerical,hillier2012numerical2,Jenkins:2022,Donne:2024}. The transition to turbulence in prominence simulations is marked by the emergence of secondary instabilities, such as the KH instability, which facilitates the turbulent energy cascade \citep[previously explored in observations][]{leonardis2012turbulent,hillier2017investigating}. In our simulation, the interaction of gravity and buoyancy induces shear flows at the edges of vertically dominated flows, leading to the presence of complex motions including KH instability-driven shear flows as shown in Figures~\ref{khi_ts}~\&~\ref{khi_stats}. As turbulence further develops and fine-scale structures form through secondary processes, short-lived jet exhausts are found within our simulation, shown in Figure~\ref{jets}, indicating that these shear flows distort the magnetic field and promote the formation of current sheets and magnetic reconnection \cite{hillier2021jets}. In our simulations, the formation of these infrequent jets facilitates short lived energy transfer and release in the upper solar atmosphere, where the plasma~$\beta$ is significantly higher than in lower regions (as shown in MC23, Figure~3). 
    
    The time and lengthscales of the secondary instabilities recovered from our simulation appear to agree with those derived from observations \citep[][]{2010ApJ...716.1288B,2018ApJ...864L..10H, Yang2018}. As shown in Figure~\ref{khi_stats}, however, the majority of the structures or dynamics of interest are present in the hotter coronal or intermediate PCTR environments, rather than the cooler prominence material. Given that these structures have been predominantly recorded in observations of prominences in the cooler, chromospheric and transition region lines, this places our results in direct contrast with previous results from observations. This is further emphasised in the following synthetic analysis, after which we will suggest potential extensions to the base model to address these discrepancies.

    The results presented in our paper correspond to a very high-resolution resistive MHD study. 
    In section~\ref{ss:synthetic}, we presented the synthetic observation equivalents to the simulation. Herein, all of the considered passbands described clear signatures of the `cells' demarcating the discrete bubbles containing rotational motions \citep[as found in the observations of e.g.,][]{2017ApJ...850...60B,Rees-Crockford:2024,Zhang:2024}. In general, the appearance of the prominence in the 171 and 193~\AA\ passbands are qualitatively similar, in contrast to the 094~\AA\ passband where the prominence absorption is expectedly weak. The jet feature of Section~\ref{sss:jets} is, nevertheless, well captured in the latter. However, when convolving these synthetic observations at a resolution of 0.02$''$ with the PSF and 0.6$''$ platescale of the AIA instrument, all of the passbands considered lose a significant amount of the finescale detail. Indeed, Figure~\ref{AIAres_171} demonstrates that the resolution of the SDO/AIA instruments is insufficient to recover the smallest scales, yet the larger cells with radii of several Mm remain visible.
    Based on these syntheses, it appears that particularly bright, curvilinear intra-prominence sheets between the dark falling fingers are strongly correlated to locations of shear flows and/or current sheets.
    Geometric considerations between this 2.5D representation at that of the 3D Sun will complicate any direct comparison.
    Nevertheless, given the much higher spatial $\approx 100$~km perihelion resolution of Solar Orbiter's \textit{Extreme Ultraviolet Imager} \citep[EUI;][]{Rochus:2020aap} \textit{High Resolution Telescope} (HRT) instrument, instances of this should be identifiable and verifiable \citep[cf.][]{Jenkins:2022}.
    
    The approximately linear dependency of the hydrogen H$\alpha$ absorption coefficient $\alpha_\lambda$ on pressure and density \citep[][]{2015A&A...579A..16H} leads the associated synthesis to match the primitive distributions almost 1\,--\,1, see Figure~\ref{density_compressibility}. Consequently, the roll-ups associated with the collision between the falling fingers and the underlying chromosphere is well captured. Once again, the geometry of the simulation at hand, the definition of the chromosphere, and the adopted magnetic field will influence how such an evolution would manifest within the 3D atmosphere. Despite these limitations, these results suggest a serendipitous observation may indeed be able to capture such an evolution if the material is not dispersed before reaching chromospheric heights \citep[as also suggested by][]{2018ApJ...869..136K}.
    
    \textcolor{black}{The motivating observations of \citet{hillier2021jets} were recorded using IRIS, whose slit-jaw imager (SJI) has a spatial resolution higher than SDO/AIA but still 10$\times$ lower than the present simulation (MC23). The spatial extent of the simulated jet is once more too small to be directly imaged. More fundamentally, however, the formation temperature of either the Mg~{\sc ii} or Si~{\sc iv} lines observed with SJI are several orders of magnitude lower (10$^{4\,-\,5}$~K) than recorded in SDO/AIA 094~\AA\ (10$^{6\,-\,7}$~K). Hence, although a jet was self-consistently formed within our simulated solar prominence undergoing a combination of nonlinear RT and KH instabilities, it is clearly not of the same origin as those that were observed by \citet{hillier2021jets}. Indeed, it is also completely absent from the cooler hydrogen H$\alpha$ dynamics. To first order, the location of the jet with respect to dense plasma is responsible for the difference in appearance, that is, the jet occurred within more `coronal' plasma already at $>$10$^6$~K. Should reconnection occur internal to a cool, dense plasma concentration then, even though the reconnection site is almost certainly too small to be directly imaged, the subsequent heating may be observed. 
    Similar to how it appears the passage of shocks within the simulation are visible within the hydrogen H$\alpha$ syntheses. Recently, \citet{Jercic:2024} performed 1.5D non-local thermodynamic equilibrium (NLTE) spectral synthesis to demonstrate that a reconnection event internal to a cool $\sim$~10$^4$~K prominence condensation yields a significant increase in the emission of the Mg~{\sc ii} line core.}

    There remain a number of improvements that can be made to the current simulation. The 2.5D domain with a constant magnetic field such that $\vec{k}\cdot\textbf{B}\approx0$ has here enabled us to achieve a high spatial resolution, resolve the numerical resistivity, and hence reproduce RH and KH instability dynamics at those scales \citep[cf.][]{Kaneko:2015,2021A&A...646A.134J}. It is imperative, however, that we explore the linear and nonlinear developments of these instabilities, including subsequent secondary processes, under more physically-realistic conditions.
    
    Considering the thermodynamics, this current model neglects the non-adiabatic terms such as radiative losses that may lead to the formation of additional, cooler material via the aforementioned KH mixing \citep[][]{Hillier:2023}. This is of particular interest as the secondary processes we find here are located within these mixing regions. For the magnetic field, we should extend the domain to consider magnetic flux ropes. Previously, \citet{2021A&A...646A..93P} explored the linear development of the RT instability within a 2.5D domain that included magnetic shear so as to approximate the internal topology of flux ropes, but as shown in \citet{Jenkins:2022} and \citet{Donne:2024} prominences are anything but invariant in the third dimension. Both authors demonstrated clearly how plasma processes throughout a flux rope can influence the development in any given $\vec{k}\cdot\textbf{B}\approx0$ plane, as a simple consequence of condensations being localised but free to move along field lines. Furthermore, and as discussed in Section~\ref{sss:KHI}, the 2.5D description artificially suppresses any influence that the magnetic tension may have. This is certain to play a larger role in the less dense environments, currently dominating the KH unstable populations, than the more dense prominence fingers. 
    
    Finally, the syntheses presented in Section~\ref{ss:synthetic} consider a LOS perpendicular to the plane, that is, along the invariant direction. An actual observation contains structure in this direction and would invariably alter the appearance, even if the underlying plasma distribution resembled closely our simulation \citep[cf.][]{Gunar:2018}. Whether for the underlying physics or the subsequent observational synthesis and as similarly concluded in MC23, the computationally advantageous yet nevertheless simplified 2.5D domain setup commonly employed to study the internal dynamics of prominences is fast becoming insufficient. We are already actively engaged in a 3D extension to this work.

\section{Conclusions} \label{s:conclusions}
In this paper, we have used 2D high-resolution resistive MHD simulations to analyse the spatio-temporal dynamics of a quiescent prominence. The system starts from an initial state where the prominence boundary is set up within the box and the RT instability is set to self-consistently evolve the system from equilibrium. The initial perturbation leads to the formation of rising plumes and falling bubbles forming coherent structures commonly present in observations of quiescent prominences. The addition of resistive dynamics leads to the formation of secondary instabilities and the growth of nonlinear dynamics which in turn excites energetic jet-like events in the prominence environment. A combination of these events forming outside of relevant temperature domains and insufficient observational resolution, comparatively, currently renders a direct 1\,--\,1 comparison impossible. Further study is necessary to map the available parameter space, with future endeavours focused on the step-up to 3D so as to excite yet-more realistic evolutions.

\begin{acknowledgements}
We acknowledge the comments and suggestions by the anonymous referee that significantly improved the accuracy, completeness, and thus relevance of this work.
This work is supported by FWO grant G0B9923N Helioskill and by Internal funds KU Leuven, through the project C16/24/010 UnderRadioSun. JMJ is supported by a European Space Agency (ESA) Internal Research Fellowship.
\end{acknowledgements}

%\bibliographystyle{aa} % style aa.bst
%\bibliography{main_bib.bib} % your references Yourfile.bib
\input{output.bbl}

\end{document}

%% file: output.bbl
\providecommand{\noopsort}[1]{}\providecommand{\singleletter}[1]{#1}%